\documentclass[12pt,journal,a4paper,draftcls,oneside,onecolumn]{IEEEtran}

\usepackage{times} 
\usepackage{graphicx}
\usepackage{cite}
\usepackage{citesort}
\usepackage{color}
\usepackage{psfrag}
\usepackage{subfigure}
\usepackage{amssymb}
\usepackage{epsfig}
\usepackage{pifont}
\usepackage{amsmath}
\usepackage{array}
\usepackage{multicol}
\usepackage[T1]{fontenc}
\usepackage[ruled, lined, linesnumbered, commentsnumbered, longend]{algorithm2e}
\usepackage{algorithmic}

\renewcommand{\baselinestretch}{1.2} 
\makeatletter
\def\ifundefined{\@ifundefined}
\makeatother 

\begin{document}


\title{{Principles and Optimization of \\Reflective Intelligent Surface Assisted mmWave Systems}
\author{Lie-Liang Yang}
\thanks{L.-L. Yang is with the School of Electronics and Computer Science, University of Southampton, SO17 1BJ, UK. (E-mail: lly@ecs.soton.ac.uk). This document is a chapter in the book: L.-L. Yang, J. Shi, K.-T. Feng, L.-H. Shen, S.-H. Wu and T.-S. Lee, Resource Optimization in Wireless Communications: Fundamentals, Algorithms and Applications, Academic Press, USA (to be published in 2024).}}


\maketitle

\begin{abstract}
A conceptual example is first analyzed to show that efficient wireless communications is possible, when user equipment (UE) receiver, BS transmitter or/and the scatter (reflector) in wireless channels employ the required channel state information (CSI) to remove the randomness of signal phase. Then, the principles and optimization of three reflective intelligent surface (RIS) assisted mmWave (RIS-mmWave) models are introduced. The first model assumes one BS, one RIS and one UE; the second one assumes one BS, one RIS and multiple UEs; while the third RIS-mmWave model assumes one BS, multiple RISs and multiple UEs. Furthermore, the optimization of BS precoder and RIS phase-shifts is addressed in the context of the massive RIS-mmWave scenarios, where the number of BS antennas and that of RIS reflection elements are significantly larger than the number of supported UEs. The analyses demonstrate that, while the deployment of RISs with mmWave is capable of solving the blockage problem and has the potential to significantly improve efficiency, finding the near-optimum solutions for RIS phase-shifts is highly challenging in practice.
\end{abstract}

\section{Introduction}

Reflection and scattering always play important roles in wireless
communications. When frequency is relatively low, a radio signal sent
by a transmitter may be reflected/scattered many times by different
objects distributed near or on the route from transmitter to receiver,
making the received signal a linear combination of the signals come
from different routes, named as propagation paths. When many signals by different  propagation paths arrive at receiver at similar time, the {\em central-limit
  theorem} can be applied to explain their
overall behavior, and the received signal can be modeled as a complex
Gaussian random variable, generating the so-called Rayleigh fading, if
there isn't a line-of-sight (LoS) or dominate path between transmitter
and receiver. Otherwise, if there is a dominate path from transmitter
to receiver, the magnitude of received signal obeys the Rician
distribution, yielding Rician
fading~\cite{book:bluebook-v2,book:Wireless-Parsons}.

As the distances of the physical paths from a transmitter to a receiver
are different, the received signals from different paths have
different delays, yielding the concept of delay-spread accounting for
the relative delay between the earliest and latest arrivals of a same
transmitted signal. When the delay spread is comparable to the symbol
duration, serious inter-symbol interference (ISI) may occur, if receiver chooses to ignore it. By
contrast, if receiver makes an effort to resolve the received signal
into several component signals with respect to the delay-spread, the
component signals can be coherently combined to obtain diversity
gain for performance improvement~\cite{Proakis-5th}.

Similarly, when transmitter, receiver or/and reflectors/scatters are
in moving, different physical paths may experience different
Doppler-frequency shifts, giving the concept of Doppler-spread
accounting for the frequency shifting range of a transmitted frequency
resulted from the above-mentioned mobility. This phenomenon can make
the orthogonally transmitted subcarrier signals become non-orthogonal,
generating inter-carrier interference (ICI). Again, if receiver is capable of resolving the
received signal into several component signals with respect to the
Doppler-spread, the component signals can be jointly exploited to
obtain diversity gain~\cite{Lie-Liang-TVT-2007-731,7925924}.

In the conventional wireless communications operated in relatively
low-frequency bands, wireless channels generated by numerous nature
reflectors/scatters are uncontrollable. When wireless communications
enters the mmWave and Terahertz eras, radio signals propagate more and
more as visible lights, there are usually only a very few of rays
(paths) from transmitter to receiver, with the LoS path carrying most
of transmit power, which however has a not ignorable probability to be
blocked. If LoS path is blocked, communication performance
significantly degrades. To provide seamless services of similar performance, in this case, alternative propagation paths are
needed. Reflective intelligent surfaces
(RISs)~\cite{9140329,9086766,9424177} have been proposed for this
purpose and, furthermore, for actively controlling radio propagation
environments to improve the performance of wireless communications.

RISs can be classified as passive RISs~\cite{9998527} and active
RISs~\cite{9998527}. A passive RIS is designed to be equipped with a large number
of reflective elements, each of which is able to reflect the incident
signal with a controllable phase shift, but is unable to control the
amplitude of the incident signal. A passive RIS does not employ active
radio-frequency components. Hence, it consumes near-zero
direct-current (DC) power, and introduces near-zero thermal noise. In
contrast, an active RIS element can reflect the incident signal with a
modified phase shift, and also scale the amplitude of the reflected
signal. Accordingly, the active processing may consume certain DC
power, and also introduce some thermal noise.

In the above-mentioned passive and active RISs, a RIS element only
reflects the signal impinging on itself. More advanced RISs may be
designed to make a RIS element emit the signals received by the
other reflection elements, in addition to reflecting the signals
impinging on itself~\cite{9737373}.

This chapter provides the principles of RIS and analyzes the
optimization in RIS-mmWave systems. Specifically, in
Section~\ref{section-6G-2.1}, a conceptual example is analyzed to show
the principles of information transmission in a wireless system, when
receiver, transmitter or/and reflector employs CSI. Then, in
Sections~\ref{section-6G-2.2.1}-\ref{section-6G-2.2.3}, three
RIS-mmWave system models are analyzed. Specifically, the first model
in Section~\ref{section-6G-2.2.1} assumes one BS, one RIS and one UE;
the second one in Section~\ref{section-6G-2.2.2} assumes one BS, one
RIS and multiple UEs; while the third RIS-mmWave model in
Section~\ref{section-6G-2.2.3} assumes one BS, multiple RISs and
multiple UEs. Some challenges in the implementation of RISs are discussed
in Section~\ref{section-6G-2.3}.

\section{A Conceptual Example}\label{section-6G-2.1}

\begin{figure}[tb]
  \begin{center}
    \includegraphics[width=0.65\linewidth]{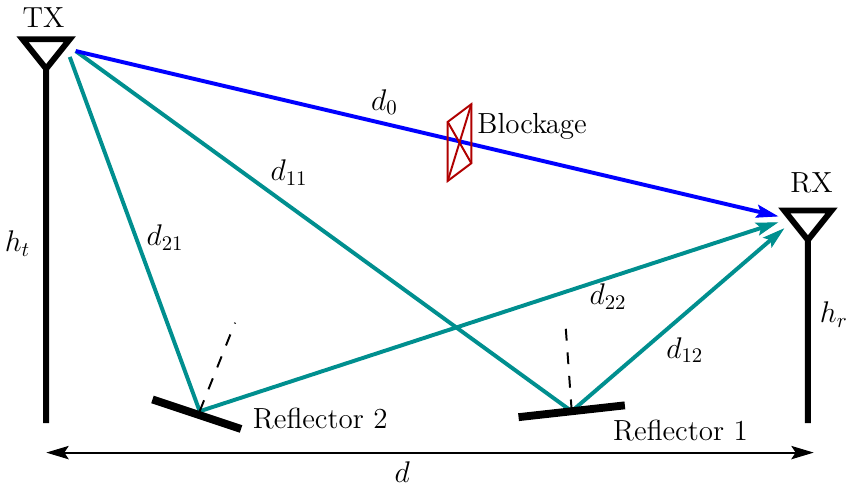}
  \end{center}  
  \caption{A simple radio propagation model.}
\label{figure-RIS-example}
\end{figure}

Following the historical development of wireless systems, in this
subsection, a simple model as shown in Fig.~\ref{figure-RIS-example}
is analyzed to show the impact of reflectors on radio signal
transmission and the possible opportunities provided for the design of
wireless systems. As shown in Fig.~\ref{figure-RIS-example}, the model
considers one transmitter, one receiver and two reflectors of
sufficiently separated. There are three transmission paths from
transmitter to receiver, one LoS path that may or may
not be blocked, and two reflected paths. The model parameters $d$ and
$h$ represent distances and heights, respectively. The objective of
design is to maximize signal-to-noise ratio (SNR) at receiver, under
the condition of the channel knowledge available to transmitter,
reflectors, or/and receiver. Note that, SNR is considered because it is the reflection of reliability, and the spectral-efficiency of a communication scheme is a monotonically increasing function of SNR.  Below several specific cases
are addressed. Except the first and second ones considering LoS path,
all the other cases assume that LoS path is blocked.

In the first case, we assume that there is only a LoS path between
transmitter and receiver, which is separated by a distance
$d_0$. Then, when a radio signal $x(t)$, satisfying $E[\|x(t)\|^2]=1$,
sent by transmitter using power $P_t$ propagates over free-space, the
received signal can be represented as
\begin{align}\label{eq:RIS-1}
r(t)=\sqrt{P_t}C_0\left(\frac{e^{j\phi_0}}{d_0}\right)x(t)+n(t)
\end{align}
where $n(t)$ is the complex Gaussian noise, assumed to be distributed
with mean zero and variance $\sigma^2/2$ per dimension, expressed as
$n(t)\sim CN(0,\sigma^2)$, and $C_0$ is a certain constant. For
example, assume that the isotropic antenna is employed. Then,
according to Friis formula for free-space
propagation~\cite{Proakis-5th}, we have
$C_0=\sqrt{G_tG_r}\lambda/4\pi$, where $G_t$ and $G_r$ are gains of
transmit and receive antennas, respectively, and $\lambda$ is
wavelength. In \eqref{eq:RIS-1}, $\phi_0$ is a phase resulted from LoS
propagation, which is dependent on radio wave's traveling distance and
wavelength, given as $\phi_0=2\pi
d_0/\lambda$~\cite{Henry-Propagation-book}. From \eqref{eq:RIS-1},
the SNR at receiver can be expressed as
\begin{align}\label{eq:RIS-2}
\gamma=C_0^2\left(\frac{1}{d_0^2}\right)\times\frac{P_t}{\sigma^2}
\end{align}
showing that the SNR for information detection decays with the square
of distance $d_0$.

The second case assumes that the received signal consists of the LoS path
and one reflected path by ground, which has been considered in many
textbooks on wireless channel
modeling~\cite{Henry-Propagation-book,book:Wireless-Parsons,book:Wireless-Rappaport-2ed}. Also,
we assume that neither transmitter nor receiver know that there are
two paths.  In this case, when $x(t)$ is transmitted, the received
signal can be expressed as
\begin{align}\label{eq:RIS-3}
r(t)=\sqrt{P_t}C_0\left(\frac{e^{j\phi_0}}{d_0}+\frac{R_1e^{j\phi_1}}{d_1}\right)x(t)+n(t)
\end{align}
where $R_1\approx -1$ is the reflection coefficient, $\phi_1=2\pi
d_1/\lambda$ with $d_1=d_{11}+d_{12}$, as shown in
Fig.~\ref{figure-RIS-example}. According to
Fig.~\ref{figure-RIS-example}, geometrically, we can obtain
$d_0=\sqrt{d^2+(h_t-h_r)^2}$, and $d_1=\sqrt{d^2+(h_t+h_r)^2}$. Assume
that $d>>h_t$, $h_r$ and also $d>>h_th_r$. We have the approximations
of $d_0=d+\frac{(h_t-h_r)^2}{2d}$ and $d_1=d+\frac{(h_t+h_r)^2}{2d}$.
Substituting these results into \eqref{eq:RIS-3} yields
\begin{align}\label{eq:RIS-4}
r(t)=&\sqrt{P_t}C_0\left(\frac{e^{j\frac{2\pi}{\lambda}\left(d+\frac{(h_t-h_r)^2}{2d}\right)}}{d+\frac{(h_t-h_r)^2}{2d}}-\frac{e^{j\frac{2\pi}{\lambda}\left(d+\frac{(h_t+h_r)^2}{2d}\right)}}{d+\frac{(h_t+h_r)^2}{2}}\right)x(t)+n(t)\nonumber\\
\approx &\sqrt{P_t}C_0\left(\frac{e^{j\frac{2\pi}{\lambda}\left(d+\frac{(h_t-h_r)^2}{2d}\right)}}{d}-\frac{e^{j\frac{2\pi}{\lambda}\left(d+\frac{(h_t+h_r)^2}{2d}\right)}}{d}\right)x(t)+n(t)\nonumber\\
=& \sqrt{P_t}C_0\left(\frac{e^{-j \frac{2\pi h_th_r}{d\lambda}}-e^{j \frac{2\pi h_th_r}{d\lambda}}}{d}\right)x(t)+n(t)\nonumber\\
=&\sqrt{P_t}C_0\left(\frac{-2j\sin\left(\frac{2\pi h_th_r}{d\lambda}\right)}{d}\right)x(t)+n(t)\nonumber\\
\approx & \sqrt{P_t}C'_0 \left(\frac{h_th_r}{d^2}\right)x(t)+n(t)
\end{align}
where $\sin(x)\approx x$ is applied, and $C_0'=-\sqrt{G_tG_r}$ if we
use $C_0=\sqrt{G_tG_r}\lambda/4\pi$. Now from \eqref{eq:RIS-4}, the
received SNR is
\begin{align}\label{eq:RIS-5}
\gamma=(C'_0)^2\left(\frac{h_t^2h_r^2}{d^4}\right)\times\frac{P_t}{\sigma^2}
\end{align}
showing that the received power decreases with the fourth power of the
distance $d$ between transmitter and receiver. Note that, as the
approximated $C_0'$ is not dependent on frequency, the SNR of
\eqref{eq:RIS-5} is less dependent on frequency than \eqref{eq:RIS-2}.

In the third case, we assume that there is no LoS path, but there are
two reflected paths from transmitter to receiver. We assume that
transmitter has no channel knowledge, reflectors are incapable of
doing anything except reflecting the incident signal, while receiver
is unable to distinguish the two incoming rays. Then, the received
signal can be represented
as~\cite{book:Wireless-Parsons,Henry-Propagation-book}
\begin{align}\label{eq:RIS-6}
r(t)=\sqrt{P_t}C_0\left(\frac{R_1 e^{j\phi_1}}{d_1}+\frac{R_2 e^{j\phi_2}}{d_2}\right)x(t)+n(t)
\end{align}
where $ 0<R_i\leq 1$, $i=1,2$, are reflection coefficient,
$\phi_i=2\pi (d_{i1}+d_{i2})/\lambda$, while $d_i$ is a function of
$d_{i1}$ and $d_{i2}$, which is dependent on the properties of
reflector. For example, according to
literature~\cite{Henry-Propagation-book,book:Wireless-Parsons,book:Wireless-Rappaport-2ed,9086766,9206044},
when in near-field transmission and the size of reflectors is large
with respect to transmission distance, $d_i=d_{i1}+d_{i2}$ is
satisfied. In contrast, when in the far-field transmission where
reflectors are small relative to transmission distance, making them
approximately as points, we have $d_i=d_{i1}d_{i2}$.

Since receiver is unable to distinguish the individual paths, it has
to treat them as one and view \eqref{eq:RIS-6} in the form of
\begin{align}\label{eq:RIS-6a}
r(t)=\sqrt{P_t}C_0\alpha(\phi_1,\phi_2) e^{j\phi(\phi_1,\phi_2)}x(t)+n(t)
\end{align}
with 
\begin{align}\label{eq:RIS-6b}
\alpha(\phi_1,\phi_2)=&\sqrt{\left(\frac{R_1 \cos(\phi_1)}{d_1}+\frac{R_2 \cos(\phi_2)}{d_2}\right)^2+\left(\frac{R_1 \sin(\phi_1)}{d_1}+\frac{R_2 \sin(\phi_2)}{d_2}\right)^2}\nonumber\\
=&\sqrt{\frac{R^2_1}{d^2_1}+\frac{R^2_2}{d^2_2}+\frac{2R_1R_2 \cos(\phi_1-\phi_2)}{d_1d_2}}\nonumber\\
\phi(\phi_1,\phi_2)=&\tan^{-1}\left(\frac{\frac{R_1 \sin(\phi_1)}{d_1}+\frac{R_2 \sin(\phi_2)}{d_2}}{\frac{R_1 \cos(\phi_1)}{d_1}+\frac{R_2 \cos(\phi_2)}{d_2}}\right)
\end{align}
To execute signal detection, receiver first estimates $\phi$ in
\eqref{eq:RIS-6a} and then carries out the detection after removing
the effect of $\phi$. The SNR of detection is
\begin{align}\label{eq:RIS-7}
\gamma=C_0^2\alpha^2(\phi_1,\phi_2)\times\frac{P_t}{\sigma^2}
\end{align}
It shows that SNR is dependent on the difference between $\phi_1$ and
$\phi_2$, and hence fluctuates. In other words, the two paths are
interfering waves, their sum can be constructive or destructive.

The fourth case uses the same assumptions as the third one, except
that the receiver is now able to distinguish the two incoming paths
and tracking the gains and phases of the two paths. In other words, the
receiver can obtain two received signals as\footnote{No interference
  between two paths is assumed.}
\begin{align}\label{eq:RIS-8a}
r_1(t)=&\sqrt{P_t}C_0\left(\frac{R_1e^{j\phi_1}}{d_1}\right)x(t)+n_1(t)\nonumber\\
r_2(t)=&\sqrt{P_t}C_0\left(\frac{R_2e^{j\phi_2}}{d_2}\right)x(t)+n_2(t)
\end{align}
and is able to obtain the knowledge about ${R_1e^{j\phi_1}}/{d_1}$ and
${R_2e^{j\phi_2}}/{d_2}$. Hence, it can implement the maximal ratio
combining (MRC)\cite{Proakis-5th}
to maximize the receive SNR\footnote{Note that MRC is optimum to
  maximize SNR.}, generating the decision variable
\begin{align}\label{eq:RIS-8}
r(t)=&\frac{R_1e^{-j\phi_1}}{d_1}r_1(t)+\frac{R_2e^{-j\phi_1}}{d_2}r_2(t)\nonumber\\
=&\sqrt{P_t}C_0\left(\frac{R_1^2}{d_1^2}+\frac{R_2^2}{d_2^2}\right)x(t)+\frac{R_1e^{-j\phi_1}}{d_1}n_1(t)+\frac{R_2e^{-j\phi_2}}{d_2}n_2(t)
\end{align}
Correspondingly, the receive SNR is
\begin{align}\label{eq:RIS-9}
\gamma={C_0^2}\left(\frac{R_1^2}{d_1^2}+\frac{R_2^2}{d^2_2}\right)\times\frac{P_t}{\sigma^2}
\end{align}
which is a constant. Hence, the achieved performance is stable for a
given $P_t/\sigma^2$.

Note that, \eqref{eq:RIS-8} explains that, to maximize receive SNR,
the MRC enhances the stronger path and discourages the weaker path in
the detection.

In the fifth case, the assumptions are the same as that of the third
case, except that the transmitter now has the knowledge about the
phases introduced by propagation, i.e., knows $\phi_1$ and
$\phi_2$. In this case, when assuming that transmitter is designed to
be able to send $e^{-j\phi_1}x(t)$ and $e^{-j\phi_2}x(t)$ over Paths 1
and 2, respectively, the received signal can be expressed as
\begin{align}\label{eq:RIS-13x1}
r(t)=\sqrt{\frac{P_t}{2}}C_0\left(\frac{R_1}{d_1}+\frac{R_2}{d_2}\right)x(t)+n(t)
\end{align}
where $P_t/2$ infers that half of power is radiated on each of the two
paths. Readily, the receive SNR in this case is
\begin{align}\label{eq:RIS-13x2}
\gamma=C^2_0\left(\frac{R_1}{d_1}+\frac{R_2}{d_2}\right)^2\times\frac{P_t}{2\sigma^2}
\end{align}

The sixth case is as the fifth case, but now the transmitter is more
capable, which can acquire full knowledge about the propagation
conditions, i.e., knows $R_i$ and $d_i$. Then, to maximize the receive
SNR, transmitter sends
\begin{align}\label{eq:RIS-10}
s_i(t)=\frac{\frac{R_ie^{-j\phi_i}}{d_i}}{\sqrt{\frac{R^2_1}{d^2_1}+\frac{R^2_2}{d^2_2}}}x(t),~i=1,2
\end{align}
via the $i$th path. Accordingly, the received signal is
\begin{align}\label{eq:RIS-11}
r(t)={\sqrt{P_t}C_0}\left(\sqrt{\frac{R^2_1}{d^2_1}+\frac{R^2_2}{d^2_2}}\right)x(t)+n(t)
\end{align}
making the receive SNR be
\begin{align}\label{eq:RIS-12}
\gamma={C_0^2}\left({\frac{R^2_1}{d^2_1}+\frac{R^2_2}{d^2_2}}\right)\times\frac{P_t}{\sigma^2}
\end{align}
It can be shown that this SNR is larger than that in
\eqref{eq:RIS-13x2}, provided that ${R_1}/{d_1}\neq {R_2}/{d_2}$.

In the last (seventh) case, the assumptions in the third case are applied,
except that the reflectors are now capable of configuring their
reflection coefficients. Specifically, when reflector $i$ is only able
to control its reflected phase by multiplying $e^{-j\phi_i}$, the
received signal is
\begin{align}\label{eq:RIS-13}
r(t)=\sqrt{P_t}C_0\left(\frac{R_1}{d_1}+\frac{R_2}{d_2}\right)x(t)+n(t)
\end{align}
Then, the receive SNR is
\begin{align}\label{eq:RIS-14}
\gamma=C^2_0\left(\frac{R_1}{d_1}+\frac{R_2}{d_2}\right)^2\times\frac{P_t}{\sigma^2}
\end{align}
By contrast, if reflector $i$ is able to control both the amplitude
and phase of its reflected signal, it can multiply $R'_i=\alpha_ie^{-j\phi_i}$,
$a_i>0$. Then, the received signal is
\begin{align}\label{eq:RIS-15}
r(t)=\sqrt{P_t}C_0\left(\frac{R_1\alpha_1}{d_1}+\frac{R_2\alpha_2}{d_2}\right)x(t)+n(t)
\end{align}
yielding the receive SNR of 
\begin{align}\label{eq:RIS-16}
\gamma=C^2_0\left(\frac{R_1\alpha_1}{d_1}+\frac{R_2\alpha_2}{d_2}\right)^2\times\frac{P_t}{\sigma^2}
\end{align}
In this case, for a given constant of $\alpha_1^2+\alpha_2^2$,
reflectors may also optimize the values of $\alpha_1$ and $\alpha_2$ to maximize the receive SNR. 

Above a few of cases have been analyzed. From these cases, we can
learn that two propagation paths may be destructive, as that in Cases
2-3, or constructive, as in Cases 4-7, for information
transmission. Specifically, Cases 1-2 demonstrate the effect of
channel on signal propagation, and Case 2 explains that an extra
ground-reflected path added on LoS path can significantly degrade the
performance of radio signal propagation. In Case 3 when receiver is
unable to distinguish individual paths, it can only treat the overall
signals as one signal to implement detection. In this case, the
receive SNR or power is varying, generating fading effect. To achieve
more efficient information transmission, the signals conveyed by the
individual paths must be added coherently at receiver. This can be
achieved when either receiver, as in Case 4, or transmitter, as in
Cases 5-6, has the state information of either the phases or both the phases
and amplitudes, of the signals conveyed by individual
paths. Alternatively, when both transmitter and receiver do not have
the state information about individual paths, but
if reflectors are capable of modifying the states of their reflected
paths, as shown in Case 7, efficient information transmission can also be
achieved. 

Notably, Cases 3-4 carry out receiver processing and, specifically,
Case 4 explains the concept of receiver diversity achieved using
MRC. The fifth and sixth cases implement transmitter preprocessing or
transmit beamforming in mmWave communications. While the fifth case
only uses channel's phase information to implement, such as, analog
beamforming, the sixth case needs knowledge of both phases and gains
about the individual channels and hence, implements digital
beamforming~\cite{19,7400949}. The last case reflects the basic
principles of relay communications in some
scenarios~\cite{1244790,5635453}, it also coincides with the
principles of RIS~\cite{9998527,9140329,9424177,9521529}. When
treated as RIS, less capable reflectors only configure their
impinging signals' phases, enabling them to be added coherently at
receiver. If more powerful reflectors are deployed, reflectors may
configure both the phases and amplitudes of the impinging signals,
so that higher SNR can be obtained at receiver and hence, to help the
information detection of higher reliability. However, it is worth
mentioning that the active processing of reflectors may introduce
noise into the signals reflected (relayed) to receiver.

In the conventional wireless systems, such as in the sub-6GHz systems that
normally experience rich scattering, communication environments or
channels are typically uncontrollable. In these systems, signal
processing can only be done at terminals with relatively powerful
signal processing capabilities, such as receiver, transmitter, relay
or their joint, after obtaining CSI with the aid of channel
estimation. By contrast, in mmWave communications, signals mainly
propagate in LoS and natural reflected signals are usually weak for
information detection. While this is a negative factor, it also
provides a method to design mmWave systems for them to deliver
information even when LoS path is blocked. Specifically, the
metasurfaces of each with many reflectors working in the principle of
that in Case 7 may be deployed to actively construct virtual LoS
(VLoS) paths from transmitter and receiver. These metasurfaces have
reconfiguration capability and hence, are referred to as reconfigurable
intelligent surfaces (RISs). 

In the following three sections, several RIS-mmWave models are
considered, from relatively simple structure to the more advanced
ones. In the following analysis, only downlink
scenarios are considered, while uplink cases can be similarly
analyzed.

\section{RIS-mmWave: Single-RIS Single User}\label{section-6G-2.2.1}

%
\begin{figure}[tb]
  \begin{center}
    \includegraphics[width=0.65\linewidth]{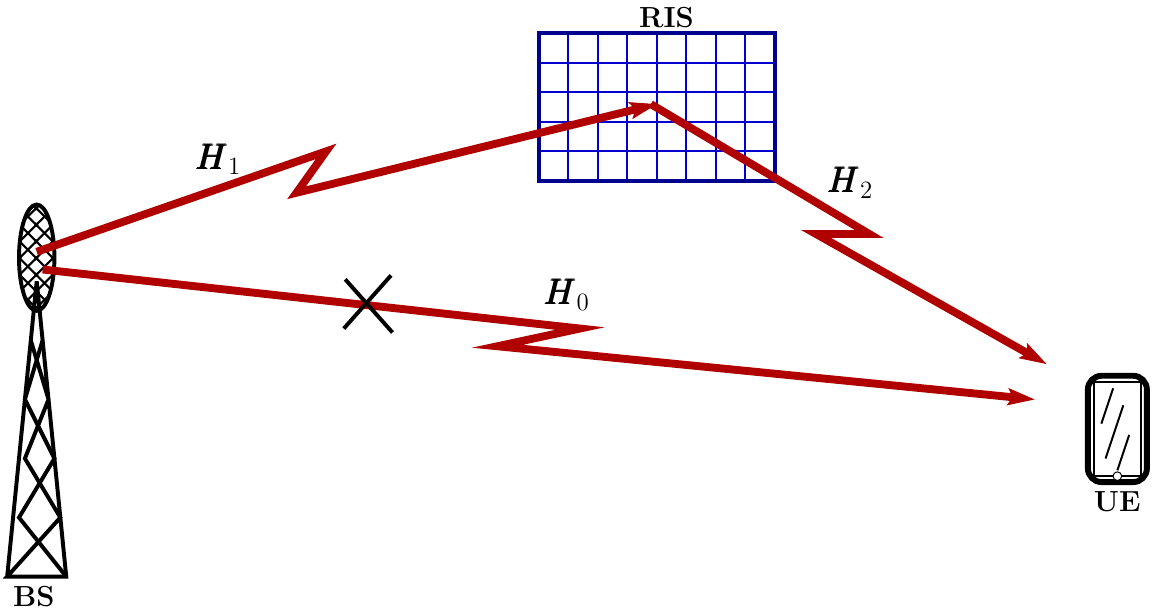} 
  \end{center}
  \caption{A RIS-mmWave system with one BS aided by one RIS to serve one user equipment.}
\label{figure-SRSU-diagram}
\end{figure}

The first RIS-mmWave model assumes that a BS, which is
aided by one RIS, serves one UE, as shown in
Fig.~\ref{figure-SRSU-diagram}. Assume that BS employs $M$ transmit
antennas and supports at most $M_d$ data streams, RIS employs $N$
reconfigurable reflecting elements, and UE has $V$ receive antennas
and $V_d$ data streams. Assume that both direct propagation link (DPL)
and reflection propagation link (RPL) between BS and UE may exist. As
shown in Fig.~\ref{figure-SRSU-diagram}, the channel transfer matrices
are $\pmb{H}_0\in\mathbb{C}^{V\times M}$ from BS to UE,
$\pmb{H}_1\in\mathbb{C}^{N\times M}$ from BS to RIS, and
$\pmb{H}_2\in\mathbb{C}^{V\times N}$ from RIS to UE.

In mmWave communications, the Saleh-Valenzuela model~\cite{1146527,9200524}
is often invoked to represent the mmWave MIMO channels. For example,
assume that uniform linear array (ULA) are employed at BS and UE,
while uniform planar array (UPA) with $N=N_x\times N_y$ is implemented
at RIS. Then, when the far-field case and narrow-band slow fading are
assumed, the channel matrices can be represented in the forms of
\begin{subequations}\label{eq:RIS-17A22}
\begin{align}
\pmb{H}_{0}=&\sum_{l=0}^{L_0}\alpha_{0,l}\pmb{a}_r\left(V,\theta^{(r)}_{0,l}\right)\pmb{a}_t^T\left(M,\theta^{(t)}_{0,l}\right)\\
\pmb{H}_{1}=&\sum_{l=0}^{L_1}\alpha_{1,l}\pmb{a}_r\left(N,\phi^{(a,r)}_{1,l},\theta^{(e,r)}_{1,l}\right)\pmb{a}_t^T\left(M,\theta^{(t)}_{1,l}\right)\\
\pmb{H}_{2}=&\sum_{l=0}^{L_2}\alpha_{2,l}\pmb{a}_r\left(V,\theta_{2,l}^{(r)}\right)\pmb{a}_t^T\left(N,\phi^{(a,t)}_{2,l},\theta^{(e,t)}_{2,l}\right)
\end{align}
\end{subequations}
In the above channel models, $l=0$ corresponds to LoS path, while the
other $L_0$, $L_1$ and $L_2$ are non-line-of-sight (NLoS) paths,
respectively, in $\pmb{H}_0$, $\pmb{H}_1$ and $\pmb{H}_2$. For the
$l$th component path, the terms in \eqref{eq:RIS-17A22} have the
meanings as follows. $\alpha_{\cdot,l}$ are the
gains. $\theta^{(t)}_{\cdot,l}$ and $\theta^{(r)}_{\cdot,l}$ represent
the angle-of-departure (AoD) and angle-of-arrival (AoA), respectively,
of ULA. $\phi^{(a,r)}_{1,l}$ and $\theta^{(e,r)}_{1,l}$ represent
the azimuth and elevation AOAs at RIS, while $\phi^{(a,t)}_{1,l}$ and
$\theta^{(e,t)}_{1,l}$ represent the azimuth and elevation AODs at
RIS. More specifically,
$\pmb{a}\left(V,\theta^{(\cdot)}_{\cdot,l}\right)$ and
$\pmb{a}\left(M,\theta^{(\cdot)}_{\cdot,l}\right)$ can be represented
in the form of~\cite{7400949,9200524}
\begin{align}\label{eq:RIS-17A23}
\pmb{a}(P,\theta)=\left[1,e^{-j2\pi\vartheta},e^{-j4\pi\vartheta},\ldots,e^{-j2(P-1)\pi\vartheta}\right]^T
\end{align}
where $P=V$ (or $M$), the normalized $\vartheta$ is obtained from the
physical angle $\theta^{(t)}_{\cdot,l}\in[-\pi/2,\pi/2]$ (or
$\theta^{(r)}_{\cdot,l}\in[-\pi/2,\pi/2]$) as
\begin{align}\label{eq:RIS-17A24}
\vartheta=\frac{d\sin(\theta)}{\lambda}
\end{align}
where $\lambda$ is wavelength and $d$ is antenna spacing. The response
vector $\pmb{a}\left(N,\phi_l,\theta_l\right)$ of RIS can be expressed
as~\cite{9898900,9200524}
\begin{align}\label{eq:RIS-17A25}
\pmb{a}\left(N,\phi_l,\theta_l\right)=\pmb{a}(N_x;\mu_{x})\otimes \pmb{a}(N_y;\mu_{y})
\end{align}
where $\otimes$ denotes the Kronecker product,
$\mu_x=\frac{d_x}{\lambda}\cos\phi_l\sin \theta_l$ and
$\mu_y=\frac{d_y}{\lambda}\sin\phi_l\sin \theta_l$, with $d_x$ and
$d_y$ denoting the antenna spacing in $x$ and $y$ directions,
respectively, and
\begin{align}\label{eq:RIS-17A26}
\pmb{a}(N_x;\mu_{x})=&\left[1~e^{-j2\pi\mu_x}~\cdots~e^{-j2\pi(N_x-1)\mu_x}\right]^T\nonumber\\
\pmb{a}(N_y;\mu_{y})=&\left[1~e^{-j2\pi\mu_y}~\cdots~e^{-j2\pi(N_y-1)\mu_y}\right]^T
\end{align}

Note that $\pmb{H}_2$ should also take into account the reflection
loss of RIS, which may be the same or different for different
reflectors.

As shown in Fig.~\ref{figure-SRSU-diagram}, the reflection matrix of
RIS is expressed as ${\pmb{\Psi}}\in\mathbb{C}^{N\times N}$, which is
typically diagonal in passive reflection, but may be non-diagonal when
advanced design is available in the future~\cite{9737373}. In
general, the elements of ${\pmb{\Psi}}$ can be expressed as
${\pmb{\Psi}}(i,j)=\Psi_{ij}e^{j\psi_{ij}}$, where $\Psi_{ij}$
represents the reflection gain (magnitude) of the signal impinging on
the $j$th element and reflected by the $i$th element, while
$\psi_{ij}$ is the phase-shift added in the impinging-reflection
process. According to definition, a passive RIS element can only
modify the phase of an incident signal, while an active RIS element
may adjust both the gain and phase of an incident signal. Hence, for
passive RIS, any element's reflection gain satisfies $0\leq
\Psi_{ij}\leq 1$, which is taken account by $\pmb{H}_2$. By contrast, for active RIS, it is possible that some elements have the gains of $\Psi_{ij}> 1$, which are provided by amplification circuitries~\cite{9998527}.

Let $\pmb{x}\in\mathbb{C}^{V_d\times 1}$, $V_d\leq M_d$,
$E[\pmb{x}]=\pmb{0}$ and $E[\|\pmb{x}\|^2]=V_d$, be a data vector sent
from BS to UE. Then, based on the above definitions, the received
signals at UE can be represented as
\begin{align}\label{eq:RIS-17}
\pmb{r}=&\left(\pmb{H}_0+\pmb{H}_2\pmb{\Psi}\pmb{H}_1\right)\pmb{F}\pmb{x}+\pmb{n}'\\
\label{eq:RIS-18}
=&\pmb{H}\pmb{F}\pmb{x}+\pmb{n}'
\end{align}
where $\pmb{H}=\pmb{H}_0+\pmb{H}_2\pmb{\Psi}\pmb{H}_1$ is the
equivalent composite channel matrix between BS and UE, $\pmb{n}'$ is the
complex Gaussian noise, each element of which is distributed with zero
mean and a variance $\sigma^2$, and $\pmb{F}\in\mathbb{C}^{M\times
  V_d}$ implements transmit beamforming, which may be a digital
beamformer or a hybrid
beamformer~\cite{7400949,HuangGuoBunton10}. In some cases, $\pmb{F}$
may be represented as
$\pmb{F}=\sqrt{\rho}\pmb{F}_D+\sqrt{1-\rho}\pmb{F}_R$ to clearly show
that there is a beam sent towards DPL and a beam sent towards RPL,
with the power assigned between two beams controlled by the parameter
$0\leq\rho\leq 1$. Correspondingly, $\rho=0$ can be viewed as a full
blockage of DPL. Note that, even when the LoS path between BS and UE
is blocked, as shown in \eqref{eq:RIS-17A22}, $\pmb{H}_0$ may contain
NLoS paths. In this case, both DPL and RPL may be at a similar
power-level, making power-allocation efficient in RIS-mmWave
systems. Otherwise, when a LoS path exists in DPL, it should dominate
the received signal of UE, making the employment of RIS less effective
or not necessary. Accordingly, all BS transmit power should be
assigned to DPL.

After UE uses a combiner $\pmb{W}\in\mathbb{C}^{V\times V_d}$, which
may also be a digital or hybrid combiner~\cite{7400949,9032163}, to
process the received signals, it obtains
\begin{align}\label{eq:RIS-19}
\pmb{y}=&\pmb{W}^H\pmb{H}\pmb{F}\pmb{x}+\pmb{n}
\end{align}
where $\pmb{n}=\pmb{W}^H\pmb{n}'$.

When both BS and UE have CSI, meaning that they know
$\{\pmb{H}_0,~\pmb{H}_1,~\pmb{H}_2\}$, and transceivers are optimally
designed for a given $\pmb{\Psi}$, the achievable
sum-rate (capacity) can be derived from \eqref{eq:RIS-18} by following the principles of
MIMO~\cite{Lie-Liang-MC-CDMA-book,MIMO-Telatar,MIMO-Telatar-I}, and
formulated as
\begin{align}\label{eq:RIS-20}
R(\pmb{\Psi})=&E_{\pmb{H}}\left\{\log_2\left[\det\left(\pmb{I}_V+\sigma^{-2}\pmb{H}\pmb{F}\pmb{Q}_x\pmb{F}^H\pmb{H}^H\right)\right]\right\}\nonumber\\
=& E_{\pmb{H}}\left\{\log_2\left[\det\left(\pmb{I}_V+\sigma^{-2}\pmb{H}\tilde{\pmb{Q}}_x\pmb{H}^H\right)\right]\right\}~\textrm{bits/s/Hz}
\end{align}
where $\pmb{Q}_x=E\left[\pmb{x}\pmb{x}^H\right]$ is the covariance
matrix of $\pmb{x}$ and $\tilde{\pmb{Q}}_x=\pmb{F}\pmb{Q}_x\pmb{F}^H$
is the covariance matrix of transmit signals. Note that, to achieve
the capacity of \eqref{eq:RIS-20}, it is the overall
$\tilde{\pmb{Q}}_x$, instead of $\pmb{Q}_x$, that is required to be optimally
designed according to $\pmb{H}$.

Based on Equation~\eqref{eq:RIS-20}, let us carry out some analysis and
gain some observations in the following two cases:
\begin{enumerate}

\item[C1:] When RIS is treated as a component of channel, the RIS-mmWave system's capacity is achieved via the joint design of transmitter and receiver. 

\item[C2:] When RIS is considered as a part of the transceiver, such as, when RIS is fully controlled by BS via a dedicated channel, the RIS-mmWave system's capacity is achieved via the joint design of transmitter, RIS and receiver.

\end{enumerate}
Specifically, in the case of C1, we may have the following observations:
\begin{itemize}
\item Given a RIS phase-shift matrix $\pmb{\Psi}$, the system is only
  required to estimate $\pmb{H}_0$ and the composite channel
  $\pmb{H}_2\pmb{\Psi}\pmb{H}_1$ to achieve optimum design.

\item Given the estimated $\pmb{H}_0$ and composite channel
  $\pmb{H}_2\pmb{\Psi}\pmb{H}_1$, the optimum design includes
  $\tilde{\pmb{Q}}_x$ and the corresponding combiner $\pmb{W}$.

\item The optimum design of $\pmb{\Psi}$ is dependent on the
  statistics of $\pmb{H}_0$, $\pmb{H}_1$ and $\pmb{H}_2$.

\item When $\pmb{\Psi}$ is fixed, $\pmb{H}$ is then fixed, and the
  RIS-mmWave system is reduced to a conventional MIMO system, whose
  optimal transceiver design is
  well-known~\cite{MIMO-Telatar,MIMO-Telatar-I}. More specifically,
  the capacity can be achieved via the singular-value decomposition
  (SVD) of $\pmb{H}$
  supported by the water-filling assisted power-allocation derived
  based on the singular values of $\pmb{H}$.  Furthermore, there are
  numerous transceiver optimization algorithms, which provide the
  design alternatives that can attain various trade-off between
  performance (efficiency) and complexity.

\item The capacity of MIMO system is dependent on the number of
  non-zero singular values, as well as the values and distributions of
  the non-zero singular values. Hence, in RIS-mmWave systems,
  $\pmb{\Psi}$ should be optimized to satisfy the requirements
  for achieving the capacity after the water-filling power-allocation.

\end{itemize}   

In comparison with the case of C1, the optimization and design of the
RIS-mmWave system in the case of C2 are more demanding. Below are some observations:
\begin{itemize}

\item System is required to estimate the individual channels of
  $\pmb{H}_0$, $\pmb{H}_1$ and $\pmb{H}_2$, which is highly
  challenging and resource-greedy, and generates large delay for a RIS-mmWave system employing a large RIS.

\item Given the estimated channels, optimum transceiver design
  includes $\pmb{\Psi}$, $\tilde{\pmb{Q}}_x$ and the corresponding
  combiner $\pmb{W}$. The optimization of $\pmb{\Psi}$ and
  $\tilde{\pmb{Q}}_x$ is coupled. Hence, they need to be jointly
  optimized to achieve capacity.

\item The optimization of $\pmb{\Psi}$ is dependent on $\pmb{H}_0$,
  $\pmb{H}_1$ and $\pmb{H}_2$.

\item Given $\pmb{H}_0$, $\pmb{H}_1$ and $\pmb{H}_2$, $\pmb{\Psi}$ may
  be optimized to make the resulted $\pmb{H}$ satisfy the
  required properties, as above-mentioned, for maximizing the
  sum-rate of system.

\end{itemize}

Explicitly, the design for achieving the capacity of RIS-mmWave
systems in either of the above-mentioned cases is high complexity. Hence,
the design of relatively low-complexity algorithms for operation in
RIS-mmWave systems has drawn the main attention in research, as seen
in \cite{9690480,9424177} and the references there in\footnote{The
  models considered in literature are usually the simplified versions
  of the RIS-mmWave model shown in Fig.~\ref{figure-SRSU-diagram}, by
  assuming, such as, single transmit antenna, single receive antenna,
  DPL and RPL contain only one LoS path, blocked DPL,
  etc.}. Typically, in a RIS-mmWave model as considered, the
phase-shifts of RIS, transmit beamformer and receive combiner can be
optimized iteratively to reduce complexity. Furthermore, instead of
working directly on the capacity of a RIS-mmWave system\footnote{To
  the best of our knowledge, the capacity of the RIS-mmWave MIMO
  systems, as that shown in Fig.~\ref{figure-SRSU-diagram}, is still
  unknown at the time of writing the book.}, its sum-rate is typically considered. For this purpose,
\eqref{eq:RIS-19} is divided into $V_d$ observations corresponding to
the $V_d$ transmitted symbols as
\begin{align}\label{eq:RIS-31}
y_s=&\pmb{w}_s^H\pmb{H}\pmb{F}\pmb{x}+\pmb{w}_s^H\pmb{H}_2\pmb{\Psi}\pmb{n}_r+n_s,~s=1,2,\ldots,V_d
\end{align}
where $\pmb{w}_s$ is the $s$-th column of $\pmb{W}$,
$n_s=\pmb{w}_s^H\pmb{n}'$. Furthermore, for the sake of generality, a
noise term $\pmb{n}_r$ is added to account for the noise possibly introduced by
RIS, which is assumed to obey the distribution
$\mathcal{CN}(\pmb{0},\sigma_r^2\pmb{I}_N)$.  Let
$\pmb{F}=\left[\pmb{f}_1,\ldots,\pmb{f}_{V_d}\right]$. Then,
\eqref{eq:RIS-31} can also be written as
\begin{align}\label{eq:RIS-32}
y_s=&\pmb{w}_s^H\pmb{H}\pmb{f}_sx_s+\sum_{i\neq s}\pmb{w}_s^H\pmb{H}\pmb{f}_ix_i+\pmb{w}_s^H\pmb{H}_2\pmb{\Psi}\pmb{n}_r+n_s
\end{align}
where, at the righthand side, the first term is the desired signal,
the second term is interference, and the other terms are Gaussian
noise. From \eqref{eq:RIS-32}, the SINR can be represented as
\begin{align}\label{eq:RIS-33}
\gamma_s=\frac{\|\pmb{w}_s^H\pmb{H}\pmb{f}_s\|^2}{\sum_{i\neq s}\|\pmb{w}_s^H\pmb{H}\pmb{f}_i\|^2+\|\pmb{w}_s^H\pmb{H}_2\pmb{\Psi}\|^2\sigma_r^2+\sigma^2},~s=1,2,\ldots,V_d
\end{align}
Accordingly, the sun-rate is given by
\begin{align}\label{eq:RIS-34}
R=\sum_{s=1}^{V_d}\log_2\left(1+\gamma_s\right)
\end{align}
which is conditioned on $\pmb{H}_0$, $\pmb{H}_1$ and $\pmb{H}_2$, as
well as $\pmb{F}$ and $\pmb{\Psi}$ designed based on the channels. The
sun-rate of \eqref{eq:RIS-34} can be directly employed as the
objective function for optimization. Furthermore, when the total power $P_t$
consumed for achieving this sum-rate is formulated, the
optimization objective function can also be defined in terms of the
energy-efficiency, which is given by the ratio of $R/P_t$~\cite{10229234}.

According to the above analysis, the general optimization problem to
maximize the sum-rate of \eqref{eq:RIS-34} includes finding the
solutions for $\pmb{\Psi}$, $\pmb{F}$ and $\pmb{W}$, under certain
constraints. However, jointly optimizing these variables for the
optimum solutions is practically impossible. Hence, the practically
meaningful algorithms are desired, which usually optimize these
variables in iterative way to obtain the solutions of approximately
optimum. Below a few of simplified models are considered.

The simplest model assumes a BS with one transmit antenna, a UE with
one receive antenna, fully blocked DPL, one passive RIS with $N$
reflection elements, only LoS path between BS and RIS and LoS path
between RIS and UE. In this simplified model,
$\pmb{\Psi}=\textrm{diag}\{e^{j\psi_1},
e^{j\psi_2},\cdots,e^{j\psi_N} \}$. Accordingly, the received signal at UE
can be written as
\begin{align}\label{eq:RIS-35}
y=&\pmb{h}^T_2\pmb{\Psi}\pmb{h}_1x+n\nonumber\\
=&|h_1||h_2|\pmb{a}_t^T\left(N,\phi^{(a,t)}_{2,0},\theta^{(e,t)}_{2,0}\right)\pmb{\Psi}\pmb{a}_r\left(N,\phi^{(a,r)}_{1,0},\theta^{(e,r)}_{1,0}\right)x+n
\end{align}  
where $\pmb{a}\left(N,\cdot,\cdot\right)$ is defined by \eqref{eq:RIS-17A25} or \eqref{eq:RIS-17A26}, $|h_1|$ and $|h_2|$ are gains of the BS-RIS and RIS-UE links\footnote{Note that, more explicitly, $|h_2|$ should be replaced by $R|h_2|$, where $R$ is the common reflection
coefficient of RIS elements, which is uncontrollable by the optimization process.}. Express the $i$th element of $\pmb{a}_t\left(N,\phi^{(a,t)}_{2,0},\theta^{(e,t)}_{2,0}\right)$ as $\pmb{a}_t(N,i)=e^{{j\varphi_{t,i}}}$, and of $\pmb{a}_r\left(N,\phi^{(a,r)}_{1,0},\theta^{(e,r)}_{1,0}\right)$ as $\pmb{a}_r(N,i)=e^{j\varphi_{r,i}}$. Then, \eqref{eq:RIS-35} can be written as
\begin{align}\label{eq:RIS-36}
y=&\left(|h_1||h_2|\sum_{i=1}^Ne^{j(\varphi_{r,i}+\varphi_{t,i}+\psi_i)}\right)x+n
\end{align}  
Hence, to make the signals be coherently added together at UE, the
phases of RIS must be set as
\begin{align}\label{eq:RIS-37}
\psi_i=-\varphi_{r,i}-\varphi_{t,i},~i=1,2,\ldots,N
\end{align}
yielding the receive SNR
\begin{align}\label{eq:RIS-38}
\gamma_s=\frac{|h_1|^2|h_2|^2N^2}{\sigma^2}
\end{align}

Equation \eqref{eq:RIS-38} infers that the attainable SNR at UE scales
with the square of the number of RIS elements, i.e., $N^2$. However,
according to \cite{9184098}, this is only valid in far-field
scenario, when RIS is physically small relative to the distances
between BS and RIS and between RIS and UE. Otherwise, when operated in the
near-field scenario, where RIS' physical size is comparable to the
BS-RIS and RIS-UE distances, the SNR only scales with $N$ of the
number of RIS elements.

In addition to the assumptions made in the above simplest model, now we assume that there also exists a path between BS and UE, i.e. DPL, which is not necessary the LoS path. Then, corresponding to \eqref{eq:RIS-36}, the model can be described as    
\begin{align}\label{eq:RIS-39}
y=&\left(|h_0|e^{j\varphi_0}+|h_1||h_2|\sum_{i=1}^Ne^{j(\varphi_{r,i}+\varphi_{t,i}+\psi_i)}\right)x+n\nonumber\\
=&e^{j\varphi_0}\left(|h_0|+|h_1||h_2|\sum_{i=1}^Ne^{j(\varphi_{r,i}+\varphi_{t,i}+\psi_i-\varphi_0)}\right)x+n
\end{align}  
Explicitly, the phases of RIS elements need to be set as 
\begin{align}\label{eq:RIS-40}
\psi_i=\varphi_0-\varphi_{r,i}-\varphi_{t,i},~i=1,2,\ldots,N
\end{align}
yielding the receive SNR
\begin{align}\label{eq:RIS-41}
\gamma_s=\frac{(|h_0|+|h_1||h_2|N)^2}{\sigma^2}
\end{align}

Next, let us consider the model where BS has $M$ transmit antennas,
passive RIS has $N$ elements, UE has one receive antenna, only RPL
exists and there is only one path from BS to RIS and one path from RIS
to UE. Accordingly, the observation obtained at UE can be written as
\begin{align}\label{eq:RIS-43}
y=&\pmb{h}^T_2\pmb{\Psi}\pmb{H}_1\pmb{f}x+n\\
\label{eq:RIS-44}
=&|h_1||h_2|\pmb{a}_t^T\left(N,\phi^{(a,t)}_{2,0},\theta^{(e,t)}_{2,0}\right)\pmb{\Psi}\pmb{a}_r\left(N,\phi^{(a,r)}_{1,0},\theta^{(e,r)}_{1,0}\right)
\pmb{a}_t^T(M,\theta_1^{(t)})\pmb{f}x+n
\end{align}  
where $|h_1|$ and $|h_2|$ are channel gains, with $|h_2|$ also accounting for the reflection loss, and $\pmb{f}$ is the
beamforming vector used by BS for transmission to RIS.

According to \eqref{eq:RIS-43}, when $\pmb{\Psi}$ is given,
$\pmb{f}$ can be optimized as
\begin{align}\label{eq:RIS-45}
\pmb{f}=\frac{1}{\sqrt{\|\pmb{h}^T_2\pmb{\Psi}\pmb{H}_1\|^2}}(\pmb{h}^T_2\pmb{\Psi}\pmb{H}_1)^H=\frac{1}{\sqrt{\|\pmb{h}^T_2\pmb{\Psi}\pmb{H}_1\|^2}}\pmb{H}_1^H\pmb{\Psi}^H\pmb{h}^*_2
\end{align}
to maximize the receive SNR at UE. Alternatively, using the
definitions below \eqref{eq:RIS-35}, and also defining
$\pmb{f}=[f_{1},f_{2},\ldots,f_{M}]^T$ and the $m$th element of
$\pmb{a}_t(M,\theta_1^{(t)})$ as $\pmb{a}_t(M,m)=e^{j\vartheta_m}$,
\eqref{eq:RIS-44} can be represented as
\begin{align}\label{eq:RIS-46}
y=&\left(|h_1||h_2|\sum_{i=1}^Ne^{j(\varphi_{r,i}+\varphi_{t,i}+\psi_i)}\times\sum_{m=1}^Mf_{m}e^{j\vartheta_m}\right)x+n
\end{align}  
Explicitly, the design of
\begin{align}\label{eq:RIS-47}
\psi_i=&-\varphi_{r,i}-\varphi_{t,i}\nonumber\\
\pmb{f}=&\frac{1}{\sqrt{M}}\pmb{a}_t^*(M,\theta_1^{(t)})
\end{align}
maximizes receive SNR, yielding
\begin{align}\label{eq:RIS-48}
y=&\left(|h_1||h_2|N\sqrt{M}\right)x+n
\end{align}
and SNR of $\gamma_s=|h_1|^2|h_2|^2N^2M/\sigma^2$, showing that the
achievable SNR also linearly scales with the number of elements of BS
antenna array.

\begin{figure}[tb]
  \begin{center}
    \includegraphics[width=0.6\linewidth]{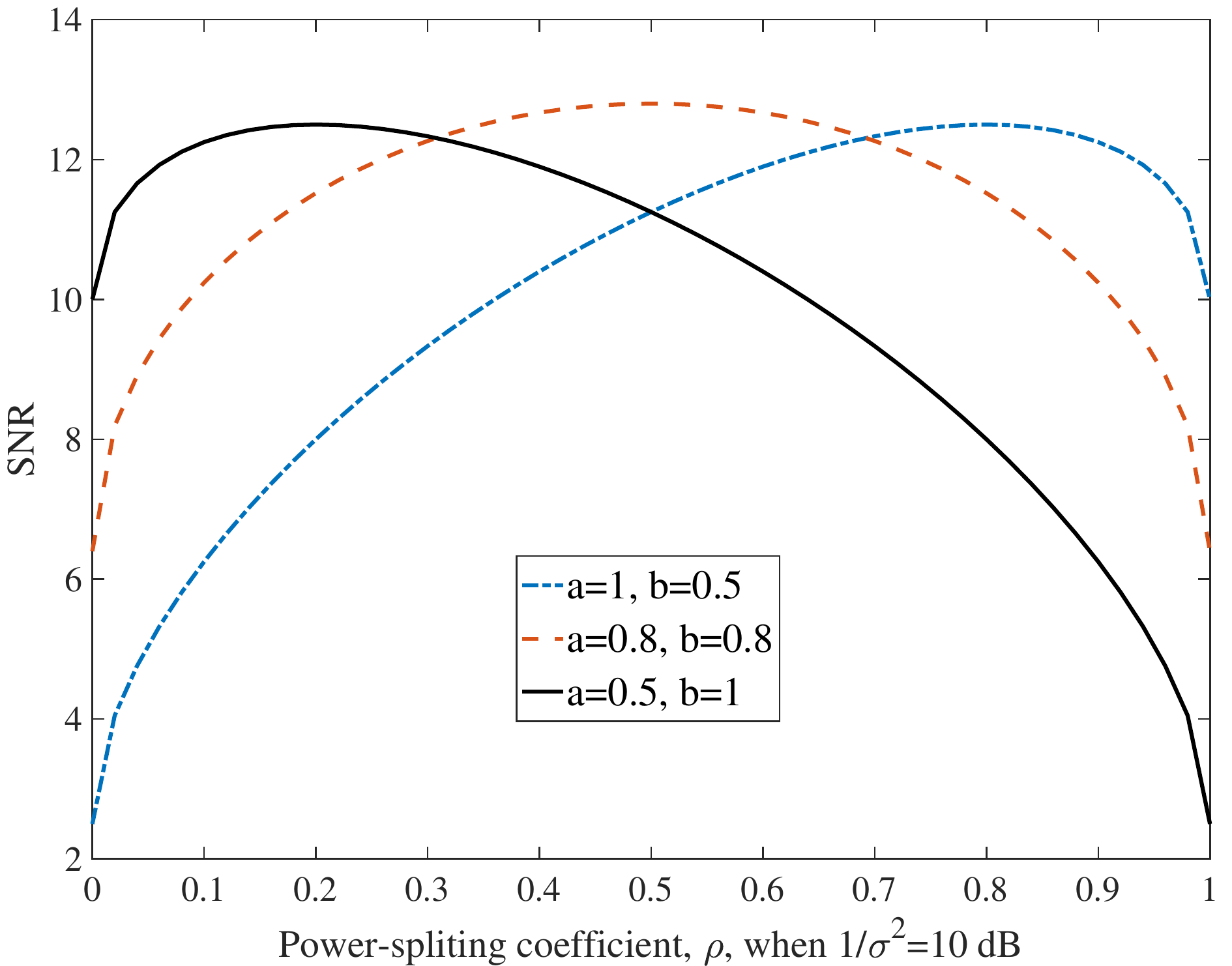} 
  \end{center}
  \caption{Achievable SNR at UE as a function of the power-allocation
    coefficient $\rho$.}
\label{figure-RIS_DRlinks_1N1}
\end{figure}

Following the above model where BS is equipped with an array but no
DPL, now we assume that there is a path directly from BS to
UE. Furthermore, to explicitly demonstrate the effect of
power-allocation between DPL and RPL, we set
$\pmb{f}=\sqrt{\rho}\pmb{f}_D+\sqrt{1-\rho}\pmb{f}_R$, where
$\pmb{f}_D$ forms a beam towards DPL and $\pmb{f}_R$ forms a beam
towards RPL. Furthermore, for simplicity, we assume that these two
beams are orthogonal. Under these assumptions, the observation at UE
can be represented as
\begin{align}\label{eq:RIS-49}
y=&\left(\sqrt{\rho}\pmb{h}_0^T\pmb{f}_D+\sqrt{1-\rho}\pmb{h}^T_2\pmb{\Psi}\pmb{H}_1\pmb{f}_R\right)x+n
\end{align}
From \eqref{eq:RIS-49} we can be inferred that the optimum design to
maximize SNR requires to jointly design $\pmb{f}_D$, $\pmb{f}_R$,
$\pmb{\Psi}$ and $\rho$. Below we consider a sub-optimum method, which
designs the precoders and RIS' phase shifts as
\begin{subequations}
\begin{align}\label{eq:RIS-50}
\pmb{f}_D=&\frac{1}{\sqrt{M}}\pmb{a}_t^*(M,\theta_0^{(t)})\\
\psi_i=&\varphi_{0}-\varphi_{r,i}-\varphi_{t,i},~i=1,2,\ldots,N\\
\pmb{f}_R=&\frac{1}{\sqrt{M}}\pmb{a}_t^*(M,\theta_1^{(t)})
\end{align}
\end{subequations}
where $\varphi_0$ is due to the representation of
$\pmb{h}_0=|h_0|e^{j\varphi_0}\pmb{a}_t^*(M,\theta_0^{(t)})$. Substituting
these terms into \eqref{eq:RIS-49} yields 
\begin{align}\label{eq:RIS-51}
y=&\left(\sqrt{\rho}|h_0|\sqrt{M}+\sqrt{1-\rho}|h_1||h_2|N\sqrt{M}\right)x+n
\end{align}
Accordingly, the receive SNR is
\begin{align}\label{eq:RIS-52}
\gamma_s(\rho)=&\frac{M\left(\sqrt{\rho}|h_0|+\sqrt{1-\rho}|h_1||h_2|N\right)^2}{\sigma^2}
\end{align}
which is a function of the power-allocation coefficient $\rho$ between
DPL and RPL.  As shown in Fig.~\ref{figure-RIS_DRlinks_1N1},
$\gamma_s(\rho)$ is a concave function of $\rho$, which has one
maximum at 
\begin{align}\label{eq:RIS-42}
\rho=\frac{a^2}{a^2+b^2}
\end{align}
with the definitions of $a=|h_0|$ and $b=|h_1||h_2|N$. Applying this $\rho$ into \eqref{eq:RIS-52} gives the maximum SNR of
\begin{align}\label{eq:RIS-52f}
\gamma_s^{\max}=&\frac{M\left(|h_0|^2+|h_1|^2|h_2|^2N^2\right)}{\sigma^2}
\end{align}

Equation \eqref{eq:RIS-42} infers, also as seen in Fig.~\ref{figure-RIS_DRlinks_1N1}, that if $a>>b$, meaning that DPL is dominant, most BS transmit power should be assigned to DPL to enable the maximum SNR. By contract, if $a<<b$, i.e., RPL is dominant, most BS power should be allocated to RPL to maximize SNR. If $a=b$, assigning half of BS power to both DPL and RPL maximizes the receive SNR. In fact, the power-allocation of \eqref{eq:RIS-42} implements the maximal ratio combining (MRC)~\cite{Proakis-5th} of the signals received by UE from DPL and RPL.

Furthermore, \eqref{eq:RIS-52} and \eqref{eq:RIS-42} infer that the
antenna array employed by BS does not affect the power-allocation
between DPL and RPL. The power-allocation is only relied on the gains
of DPL and RPL.

In contrast to the above models of all assuming passive RIS, finally,
following \cite{9998527}, we consider an active RIS that introduces reflective noise. The system is also under the constraints of both
the total transmit power $P_{B}$ at BS and the total transmit power
$P_R$ at RIS. Assume that BS is capable of attaining all the channel
knowledge of $\pmb{H}_0$, $\pmb{H}_1$ and $\pmb{H}_2$, and UE knows
$\pmb{H}_2$. In this case, UE can simply design a $\pmb{W}$ to
maximize the rate between RIS and UE based on $\pmb{H}_2$ by assuming
the virtual observations of
$\pmb{y}_2=\pmb{W}^H\pmb{H}_2\pmb{F}_2\pmb{x}_2+\pmb{n}_2$, where
$\pmb{F}_2\in\mathbb{C}^{N\times V_d}$ is a designed virtual
beamformer for RIS to send information
$\pmb{x}_2\in\mathbb{C}^{V_d\times 1}$ to UE. As BS knows $\pmb{H}_0$,
$\pmb{H}_1$ and $\pmb{H}_2$, it can also design the same $\pmb{W}$ as
UE. Hence, it can optimize its precoder $\pmb{F}$ in zero-forcing (ZF)
principle~\cite{Lie-Liang-MC-CDMA-book} to fully remove the
interference between the received symbols at UE. Consequently,
\eqref{eq:RIS-33} can be written as
\begin{align}\label{eq:RIS-53}
\gamma_s=\frac{\|\pmb{w}_s^H\pmb{H}\pmb{f}_s\|^2}{\|\pmb{w}_s^H\pmb{H}_2\pmb{\Psi}\|^2\sigma_r^2+\sigma^2},~s=1,2,\ldots,V_d
\end{align}

From \eqref{eq:RIS-19} and \eqref{eq:RIS-31}, the transmit power of BS
and RIS can be derived, which can be expressed as
\begin{subequations}
\begin{align}\label{eq:RIS-54}
P_B=&E\left[\|\pmb{F}\pmb{x}\|^2\right]=\sum_{s=1}^{V_d}\|\pmb{f}_s\|^2\\
P_R=&E\left[\|\pmb{\Psi}\pmb{H}_1\pmb{F}\pmb{x}+\pmb{\Psi}\pmb{n}_r\|^2\right]=\sum_{s=1}^{V_d}\|\pmb{\Psi}\pmb{H}_1\pmb{f}_s\|^2+\|\pmb{\Psi}\|^2_F\sigma_r^2
\end{align}
\end{subequations}
respectively. Hence, the optimization problem for designing $\pmb{F}$
and $\pmb{\Psi}$ to maximize sum-rate can be described as
\begin{subequations}
\begin{align}\label{eq:RIS-55}
\{\pmb{F}^{\textrm{opt}},\pmb{\Psi}^{\textrm{opt}}\}&=\arg\max_{\pmb{F},\pmb{\Psi}}\left\{R(\pmb{F},\pmb{\Psi})=\sum_{s=1}^{V_d}\log_2(1+\gamma_s)\right\}\\
\textrm{s.t.~~}& \textrm{C1:~} \pmb{w}_s^H\pmb{H}\pmb{f}_i=0 ~~\forall s\textrm{~and~} i\neq s\\
&\textrm{C2:~} \sum_{s=1}^{V_d}\|\pmb{f}_s\|^2\leq P_{B}^{\max}\\
&\textrm{C3:~}  \sum_{s=1}^{V_d}\|\pmb{\Psi}\pmb{H}_1\pmb{f}_s\|^2+\|\pmb{\Psi}\|^2_F\sigma_r^2\leq P_{R}^{\max}
\end{align}
\end{subequations}
where the constraint C1 forces the system to be free of ISI.  

It can be readily realized that jointly solving the above optimization
problem is extremely hard. Instead, iterative optimization of
$\pmb{F}$ and $\pmb{\Psi}$ can be executed, such as, by the methods
modified from those introduced in \cite{9998527}. Note that, in the
above optimization problem, we assume that $\pmb{W}$ has been
separately designed based only on $\pmb{H}_2$ to maximize the sum-rate
between RIS and UE.

Solving the above optimization problem has the implication of
resource-allocation. Specifically, the design needs to find the
solutions for the power assigned between DPL and RPL. Simultaneously,
for both DPL and RPL, the design needs to find the solutions to assign
corresponding power to the different beams between BS and RIS, between
RIS and UE, and between BS and UE.

\section{RIS-mmWave: Single-RIS Multiple Users}\label{section-6G-2.2.2}

%
\begin{figure}[tb]
  \begin{center}
    \includegraphics[width=0.65\linewidth]{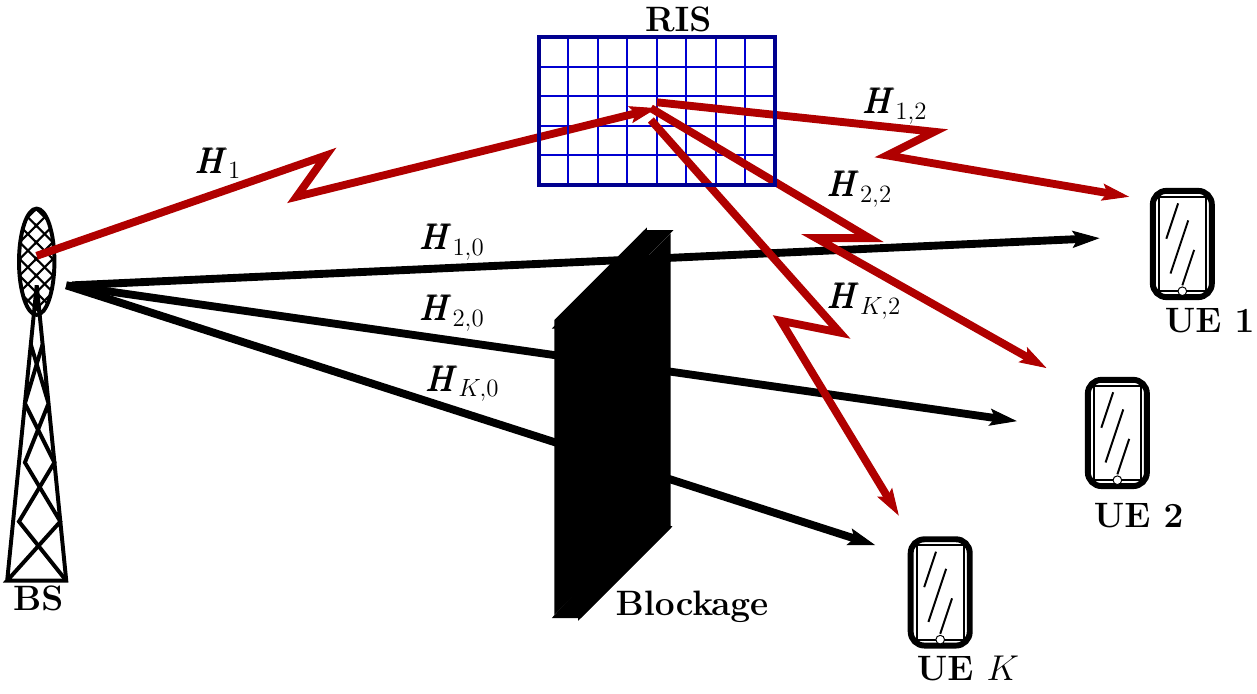} 
  \end{center}
  \caption{A RIS-mmWave system with one BS via one RIS to serve $K$ user equipments.}
\label{figure-SRMU-diagram}
\end{figure}

Above a RIS-mmWave model with single UE is considered. Now we extend
it to the scenario of multiple UEs, with the system diagram as shown in
Fig.~\ref{figure-SRMU-diagram}. We use the same settings and
assumptions as those in Section~\ref{section-6G-2.2.1}, in addition to
replacing $V$ by $V_k$, $V_d$ by $V_{k,d}$, $\pmb{H}_0$ by
$\pmb{H}_{k,0}$, $\pmb{H}_2$ by $\pmb{H}_{k,2}$, $\pmb{F}$ by
$\pmb{F}_k$, and $\pmb{W}$ by $\pmb{W}_k$. Note that, $\pmb{H}_1$ and
$\pmb{\Psi}$ are the same as them in Section~\ref{section-6G-2.2.1},
which are common to all UEs. Then, by following
Section~\ref{section-6G-2.2.1}, it can be shown that the observations
obtained by the $k$th UE can be represented as
\begin{align}\label{eq:RIS-56}
\pmb{r}_k=&\left(\pmb{H}_{k,0}+\pmb{H}_{k,2}\pmb{\Psi}\pmb{H}_1\right)\sum_{l=1}^K\pmb{F}_{l}\pmb{x}_l+\pmb{H}_{k,2}\pmb{\Psi}\pmb{n}_r+{\pmb{n}}'_k\\
\label{eq:RIS-57}
=&\sum_{l=1}^K\pmb{H}_k\pmb{F}_l\pmb{x}_l+\pmb{H}_{k,2}\pmb{\Psi}\pmb{n}_r+\pmb{n}'_k,~k=1,2,\ldots,K
\end{align}
where $\pmb{H}_k=\pmb{H}_{k,0}+\pmb{H}_{k,2}\pmb{\Psi}\pmb{H}_1$ is
the equivalent composite channel between BS and UE $k$, and
$\pmb{H}_{k,2}\pmb{\Psi}\pmb{n}_r$ is the RIS processing generated noise
forwarded to UE $k$. 

UE $k$ uses the combiner $\pmb{W}_{k}$ to form the decision
variables for detecting its $V_{k,d}$ symbols, yielding
\begin{align}\label{eq:RIS-58}
\pmb{y}_k=&\pmb{W}_k^H\pmb{r}_k\nonumber\\
=&\pmb{W}_k^H\pmb{H}_k\pmb{F}_k\pmb{x}_k+\sum_{l\neq k}\pmb{W}_k^H\pmb{H}_k\pmb{F}_l\pmb{x}_l+\pmb{W}_k^H\pmb{H}_{k,2}\pmb{\Psi}\pmb{n}_r+\pmb{n}_k,\nonumber\\
&~~k=1,2,\ldots,K
\end{align}
where in the second equation, $\pmb{n}_k=\pmb{W}_k^H\pmb{n}'_k$, the
first term corresponds to the desired UE $k$, and the second term is
multi-user interference (MUI) on
UE $k$. Depended on the BS precoder and the combiner of UE $k$, the
symbols in $\pmb{x}_k$ may or may not interfere with each other.

Corresponding to \eqref{eq:RIS-32}, the decision variable for
detecting the $s$th symbol of UE $k$ can be expressed as
\begin{align}\label{eq:RIS-59}
y_{k,s}=&\pmb{w}_{k,s}^H\pmb{H}_k\pmb{f}_{k,s}x_{k,s}+\underbrace{\sum_{i\neq s}\pmb{w}_{k,s}^H\pmb{H}_k\pmb{f}_{k,i}x_{k,i}}_{\textrm{ISI}}\nonumber\\
&+\underbrace{\sum_{l\neq k}\pmb{w}_{k,s}^H\pmb{H}_k\pmb{F}_l\pmb{x}_l}_{\textrm{MUI}}+\pmb{w}_{k,s}^H\pmb{H}_{k,2}\pmb{\Psi}\pmb{n}_r+n_{k,s}\nonumber\\
&s=1,2,\ldots,V_{k,d}
\end{align}
where
$\pmb{F}_k=\left[\pmb{f}_{k,1},\pmb{f}_{k,2},\ldots,\pmb{f}_{k,V_{k,d}}\right]$
is applied. From \eqref{eq:RIS-59} the SINR for detecting a symbol of
UE $k$ can be derived to be
\begin{align}\label{eq:RIS-60}
\gamma_{k,s}=&\frac{\|\pmb{w}_{k,s}^H\pmb{H}_k\pmb{f}_{k,s}\|^2}{\sum_{i\neq s}\|\pmb{w}_{k,s}^H\pmb{H}_k\pmb{f}_{k,i}\|^2+\sum_{l\neq k}\|\pmb{w}_{k,s}^H\pmb{H}_k\pmb{F}_l\|^2+\|\pmb{w}_{k,s}^H\pmb{H}_{k,2}\pmb{\Psi}\|^2\sigma_r^2+\sigma^2},\nonumber\\
&~s=1,2,\ldots,V_{k,d};~k=1,2,\ldots,K
\end{align}
Accordingly, the sun-rate is given by
\begin{align}\label{eq:RIS-61}
R(\pmb{W},\pmb{F},\pmb{\Psi})=&\sum_{k=1}^KR_k=\sum_{k=1}^K\sum_{s=1}^{V_{k,d}}\log_2\left(1+\gamma_{k,s}\right)
\end{align}
where by definition,
$\pmb{W}=\textrm{diag}\{\pmb{W}_1,\pmb{W}_2,\ldots,\pmb{W}_K\}$ and
$\pmb{F}=\left[\pmb{F}_1,\pmb{F}_2,\ldots,\pmb{F}_K\right]$.

From \eqref{eq:RIS-56} and \eqref{eq:RIS-57}, the transmit power of BS
and RIS can be found to be
\begin{subequations}
\begin{align}\label{eq:RIS-62}
P_B=&\sum_{k=1}^KP_{B,k}=\sum_{k=1}^K\|\pmb{F}_{k}\|^2_F\\
P_R=&\sum_{k=1}^K\|\pmb{\Psi}\pmb{H}_1\pmb{F}_{k}\|^2_F+\|\pmb{\Psi}\|_F^2\sigma_r^2
\end{align}
\end{subequations}

Then, to maximize the sum-rate of the RIS-mmWave system under the
transmit power constraint at both BS and RIS, and the constraint of
zero ISI and MUI, the optimization problem can be described as
\begin{subequations}\label{eq:RIS-63}
\begin{align}
\{\pmb{W}^{\textrm{opt}},\pmb{F}^{\textrm{opt}},\pmb{\Psi}^{\textrm{opt}}\}&=\arg\max_{\pmb{W},\pmb{F},\pmb{\Psi}}\left\{R(\pmb{W},\pmb{F},\pmb{\Psi})=\sum_{k=1}^K\sum_{s=1}^{V_{k,d}}\log_2\left(1+\gamma_{k,s}\right)\right\}\\
\textrm{s.t.~~}& \textrm{C1:~} \pmb{w}_{k,s}^H\pmb{H}_k\pmb{f}_{k,i}=0 ~~\forall k,~s\textrm{~and~} i\neq s\\
& \textrm{C2:~} \pmb{w}_{k,s}^H\pmb{H}_k\pmb{F}_{l,j}=0 ~~\forall k,~s, l,j~\textrm{~and~} l\neq k\\
&\textrm{C3:~}\sum_{k=1}^K\|\pmb{F}_{k}\|^2_F \leq P_{B}^{\max}\\
&\textrm{C4:~} \sum_{k=1}^K\|\pmb{\Psi}\pmb{H}_1\pmb{F}_{k}\|^2_F+\|\pmb{\Psi}\|_F^2\sigma_r^2 \leq P_{R}^{\max}
\end{align}
\end{subequations}
Note that in the above problem, if each data symbol is constrained to
use at most one unit of energy for its transmission, we have
$P_{B}^{\max}+P_{R}^{\max}\leq \sum_{k=1}^KV_{k,d}$.

As jointly solving the above optimization problem is extremely
difficult, iterative optimization is usually implemented. Furthermore,
considering that, in practice, the number of array elements of a UE is
usually significantly smaller than that of BS and that of RIS, little
performance gain can be attained via the iteration with respect
$\pmb{W}$. Instead, $\pmb{W}_k$ can simply be designed to maximize the
sum-rate between RIS and individual UEs, without considering the
inter-effect between UEs and the channels between BS and RIS. After
$\pmb{W}_k$ are obtained and fixed, the ISI and MUI can be tackled by
the designs of $\pmb{F}$ and $\pmb{\Psi}$. Following the principle of
massive MIMO~\cite{ref_MassiveMIMO}, when the number of array
elements at BS and the number of RIS elements are sufficiently large,
ZF-assisted precoding is near-optimum. Furthermore, ZF-assisted
precoding often provides the results that are relatively easy to analyze. Hence, the constraints of zero ISI and zero MUI are imposed, as stated in \eqref{eq:RIS-63}(b) and \eqref{eq:RIS-63}(c), on the
optimization. Note that, when $\pmb{W}$ and $\pmb{\Psi}$ are given,
the ZF precoding becomes a typical ZF transmitter preprocessing
problem in MIMO, the solution of which (including power-allocation)
can be found, for example, in \cite{Lie-Liang-MC-CDMA-book}.

\begin{algorithm}[t]
\caption{Iterative algorithm for BS to optimize $\pmb{F}$ and $\pmb{\Psi}$.}
\label{Algorithm-RIS-mmWave-SRMU}

\textbf{Inputs:} $\{\pmb{H}_{k,0}, \pmb{H}_{k,1}, \pmb{H}_{k,2}\}$ \\

\textbf{Design of Combiners:} For $k=1,2,\ldots, K$, optimize
$\{\pmb{W}_k\}$ to maximize the sum-rate of UE $k$ based on
$\{\pmb{H}_{k,2}\}$.\\

\textbf{Initialization:} Randomly set a $\pmb{\Psi}$ according to the
capability of RIS. \%\%such as active/inactive RIS, non-diagonal/diagonal  $\pmb{\Psi}$, etc. 

{
  \Repeat{Convergence achieved}{ 

1) For a given $\pmb{\Psi}$, BS optimizes $\pmb{F}$ to maximize the
sum-rate of the RIS-mmWave system with the constraints of
\eqref{eq:RIS-63}(b)-\eqref{eq:RIS-63}(e).

2) For a given $\pmb{F}$, BS optimizes $\pmb{\Psi}$ to maximize the
sum-rate of the RIS-mmWave system with or without the constraints of
\eqref{eq:RIS-63}(b) and \eqref{eq:RIS-63}(c).  } }

\end{algorithm}

With the above constraints, for example, an iterative optimization for
$\pmb{F}$ and $\pmb{\Psi}$ may be generally described as
Algorithm~\ref{Algorithm-RIS-mmWave-SRMU}. Note that the constraints of \eqref{eq:RIS-63}(b) and \eqref{eq:RIS-63}(c) may not be imposed to release the ZF precoding constraint. Furthermore, we can be inferred that the optimization
in Algorithm~\ref{Algorithm-RIS-mmWave-SRMU} may involve the
power-allocation between DPLs and RPLs, and between BS transmission
and RIS transmission, as well as the power-allocation among the data
streams of all $K$ UEs.

Below we consider a few of simplified models as well as their optimization.

The first one assumes no DPL, and that BS employs $M$ transmit
antennas to support $K$ UEs of each employing single antenna with the aid of a
passive RIS of $N$ elements. Hence, $V_{k,d}=1$,
$\pmb{\Psi}=\textrm{diag}\{e^{j\psi_1},e^{j\psi_2},\ldots,e^{j\psi_N}\}$,
and no noise added at RIS. Accordingly, the observation obtained by UE
$k$ can be written as
\begin{align}\label{eq:RIS-64}
y_{k}=&\pmb{h}_{k,2}^T\sum_{l=1}^K\pmb{\Psi}\pmb{H}_1\underbrace{\tilde{\pmb{f}}_l\sqrt{\beta_l}}_{\pmb{f}_l}x_l+n_k\nonumber\\
=&\pmb{h}_{k,2}^T\pmb{\Psi}\pmb{H}_1\underbrace{\tilde{\pmb{F}}\pmb{\beta}^{1/2}}_{\pmb{F}}\pmb{x}+n_k,~k=1,2,\ldots,K
\end{align}
where $\pmb{h}^T_{k,2}$ is the channel vector from RIS to UE $k$,
$\pmb{x}=[x_1,x_2,\ldots,x_K]^T$, and
$\pmb{F}=\tilde{\pmb{F}}\pmb{\beta}^{1/2}$ with
$\tilde{\pmb{F}}=\left[\tilde{\pmb{f}}_1,\tilde{\pmb{f}}_2,\ldots,\tilde{\pmb{f}}_K\right]$
for implementing precoding and
$\pmb{\beta}^{1/2}=\textrm{diag}\{\beta_1^{1/2},\beta_2^{1/2},\ldots,\beta_K^{1/2}\}$ for
scaling the precoder $\tilde{\pmb{F}}$ to make the transmit signal
satisfy the power constraint.

The precoder matrix $\pmb{F}$ and phase shift matrix $\pmb{\Psi}$ in
this model can be optimized using different methods. For example, by
assuming the ZF precoding, the optimization can be changed to
optimizing the power-allocation and phase shifts. In detail, let
$\pmb{y}=[y_1,y_2,\ldots,y_K]^T$, $\pmb{n}=[n_1,n_2,\ldots,n_K]^T$ and
$\pmb{H}_2=\left[\pmb{h}_{1,2},\pmb{h}_{2,2},\ldots,\pmb{h}_{K,2}\right]^T$,
which is a $(K\times N)$ matrix. Then, we have
\begin{align}\label{eq:RIS-65}
\pmb{y}=&\pmb{H}_2\pmb{\Psi}\pmb{H}_1\tilde{\pmb{F}}\pmb{\beta}^{1/2}\pmb{x}+\pmb{n}
\end{align}
Given $\pmb{H}_2$, $\pmb{\Psi}$ and $\pmb{H}_1$, and assuming that
$K\leq N$ and $K\leq M$, the ZF solution for precoding is given
by~\cite{Lie-Liang-MC-CDMA-book}
\begin{align}\label{eq:RIS-66}
\tilde{\pmb{F}}=\left(\pmb{H}_2\pmb{\Psi}\pmb{H}_1\right)^H\left(\pmb{H}_2\pmb{\Psi}\pmb{H}_1\pmb{H}_1^H\pmb{\Psi}^*\pmb{H}_2^H\right)^{-1}
\end{align}
Correspondingly, the total transmit power is
\begin{align}\label{eq:RIS-67}
P_t=&\textrm{Tr}\left(\tilde{\pmb{F}}\pmb{\beta}\tilde{\pmb{F}}^H\right)\nonumber\\
=&\textrm{Tr}\left(\pmb{\beta}\left(\pmb{H}_2\pmb{\Psi}\pmb{H}_1\pmb{H}_1^H\pmb{\Psi}^*\pmb{H}_2^H\right)^{-1}\right)
\end{align}
where $\textrm{Tr}\left(\pmb{A}\right)$ returns the trace of square
matrix $\pmb{A}$.

Substituting \eqref{eq:RIS-66} into \eqref{eq:RIS-65} yields
\begin{align}\label{eq:RIS-68}
\pmb{y}=&\pmb{\beta}^{1/2}\pmb{x}+\pmb{n}
\end{align}
Hence, the SNR of UE $k$ is
$\gamma_k(\pmb{\Psi})=\beta_k(\pmb{\Psi})/\sigma^2$, which is a
function of $\pmb{\Psi}$.  Accordingly, the sum-rate is
\begin{align}\label{eq:RIS-69}
R(\pmb{\beta},\pmb{\Psi})=\sum_{k=1}^K\log_2\left(1+\frac{\beta_k(\pmb{\Psi})}{\sigma^2}\right)
\end{align}

Consequently, constrained on the ZF precoding, the optimization
becomes to iteratively solve the problem for the power-allocation and
the phase shifts in $\pmb{\Psi}$. Specifically, during an iteration,
when $\pmb{\Psi}$ is given, the power-allocation for $\pmb{\beta}$ can
be described as
\begin{subequations}\label{eq:RIS-70}
\begin{align}
\pmb{\beta}^{opt}(\pmb{\Psi})=&\arg\max_{\pmb{\beta}}\left\{R(\pmb{\beta},\pmb{\Psi})=\sum_{k=1}^K\log_2\left(1+\frac{\beta_k(\pmb{\Psi})}{\sigma^2}\right)\right\}\\
s.t. &~~\textrm{Tr}\left(\pmb{\beta}\left(\pmb{H}_2\pmb{\Psi}\pmb{H}_1\pmb{H}_1^H\pmb{\Psi}^*\pmb{H}_2^H\right)^{-1}\right)\leq P_{B}^{\max} 
\end{align}
\end{subequations}
It can be shown that the above problem has the close-form solutions of
the water-filling style.

Note that a simpler solution~\cite{Lie-Liang-MC-CDMA-book} to
$\pmb{\beta}$ can be readily obtained from the constraint
\eqref{eq:RIS-70} (b) by letting
$\pmb{\beta}=\beta\pmb{I}_K$, yielding
\begin{align}\label{eq:RIS-71}
\beta=\frac{P_{B}^{\max}}{\textrm{Tr}\left(\left(\pmb{H}_2\pmb{\Psi}\pmb{H}_1\pmb{H}_1^H\pmb{\Psi}^*\pmb{H}_2^H\right)^{-1}\right)}
\end{align}

Furthermore, an optimization problem
that has the meaning of maximal SNR can be stated
as~\cite{Lie-Liang-MC-CDMA-book}
\begin{subequations}\label{eq:RIS-72}
\begin{align}
\pmb{\beta}^{opt}(\pmb{\Psi})=&\arg\min_{\pmb{\beta}}\left\{\sigma^2\sum_{k=1}^K\frac{1}{\beta_k(\pmb{\Psi})}\right\}\\
s.t. &~~\textrm{Tr}\left(\pmb{\beta}\left(\pmb{H}_2\pmb{\Psi}\pmb{H}_1\pmb{H}_1^H\pmb{\Psi}^*\pmb{H}_2^H\right)^{-1}\right)\leq P_{B}^{\max} 
\end{align}
\end{subequations}
which also has the closed-form solution, as detailed in\cite{Lie-Liang-MC-CDMA-book}.

On the other side, when $\pmb{\beta}$ is fixed, $\pmb{\Psi}$ can be
obtained from solving the unconstrained problem described
as~\cite{8741198}
\begin{subequations}\label{eq:RIS-73}
\begin{align}
\pmb{\Psi}^{opt}(\pmb{\beta})=&\arg\min_{\pmb{\Psi}}\left\{P_t=\textrm{Tr}\left(\pmb{\beta}\left(\pmb{H}_2\pmb{\Psi}\pmb{H}_1\pmb{H}_1^H\pmb{\Psi}^*\pmb{H}_2^H\right)^{-1}\right)\right\}\\
s.t. &~~|\pmb{\Psi}(n,n)|=1,~~\forall n=1,2,\ldots,N
\end{align}
\end{subequations}
Its solution to $\pmb{\Psi}$ can be derived with the aid of, such as,
the gradient descent approach or the sequential fractional programming
method~\cite{8741198}.

Another method introduced in \cite{9110869} is using the deep
reinforcement learning (DRL)
algorithm to directly solve the sum-rate maximization problem described as
\begin{subequations}\label{eq:RIS-73x}
\begin{align}
\{\pmb{F},\pmb{\Psi}\}=&\arg\max_{\pmb{F},\pmb{\Psi}}\left\{\sum_{k=1}^K\log(1+\gamma_k)\right\}\\
s.t.~~C1:&~~\sum_{k=1}^K\|\pmb{f}_k\|^2\leq P_B^{\max}\\
C2:&~~|\pmb{\Psi}(n,n)|=1,~~\forall n=1,2,\ldots,N
\end{align}
\end{subequations}
which motivates to jointly find the solutions to $\pmb{F}$ and $\pmb{\Psi}$, where
 the SINR in terms of UE $k$ is
\begin{align}\label{eq:RIS-73y}
\gamma_k=\frac{\left\|\pmb{h}_{k,2}^T\pmb{\Psi}\pmb{H}_1\pmb{f}_k\right\|^2}{\displaystyle\sum_{l\neq k}^K\left\|\pmb{h}_{k,2}^T\pmb{\Psi}\pmb{H}_1\pmb{f}_l\right\|^2+\sigma^2},~k=1,2,\ldots,K
\end{align}

The second simplified model extends the above one by invoking a DPL
from BS to each of UEs. Let the channel from BS to UE $k$ be expressed
as a $M$-length vector $\pmb{h}_{k,0}$. Then, corresponding to
Eqs.~\eqref{eq:RIS-64} and \eqref{eq:RIS-65}, we have
\begin{align}\label{eq:RIS-74}
y_{k}=&\left(\pmb{h}_{k,0}^T+\pmb{h}_{k,2}^T\pmb{\Psi}\pmb{H}_1\right)\pmb{F}\pmb{x}+n_k,~k=1,2,\ldots,K
\end{align}
and 
\begin{align}\label{eq:RIS-75}
\pmb{y}=&\left(\pmb{H}_0+\pmb{H}_2\pmb{\Psi}\pmb{H}_1\right)\pmb{F}\pmb{x}+\pmb{n}
\end{align}
where $\pmb{H}_0=\left[\pmb{h}_{1,0},\pmb{h}_{2,0},\ldots,\pmb{h}_{K,0}\right]^T$.

From \eqref{eq:RIS-74}, the SINR $\gamma_{k}$ can be derived to be
\begin{align}\label{eq:RIS-76}
\gamma_k=\frac{\left\|\left(\pmb{h}_{k,0}^T+\pmb{h}_{k,2}^T\pmb{\Psi}\pmb{H}_1\right)\pmb{f}_k\right\|^2}{\sum_{l\neq k}^K\left\|\left(\pmb{h}_{k,0}^T+\pmb{h}_{k,2}^T\pmb{\Psi}\pmb{H}_1\right)\pmb{f}_l\right\|^2+\sigma^2},~k=1,2,\ldots,K
\end{align}

Based on \eqref{eq:RIS-75}, it can be shown that, when $\pmb{\Psi}$ is
fixed, the ZF solution to $\tilde{\pmb{F}}$ is
\begin{align}\label{eq:RIS-77}
\tilde{\pmb{F}}=\left(\pmb{H}_0+\pmb{H}_2\pmb{\Psi}\pmb{H}_1\right)^H\left[(\pmb{H}_0+\pmb{H}_2\pmb{\Psi}\pmb{H}_1)(\pmb{H}_0+\pmb{H}_2\pmb{\Psi}\pmb{H}_1)^H\right]^{-1}
\end{align}
associated with the power constraint of
\begin{align}\label{eq:RIS-78}
P_t=&\textrm{Tr}\left(\pmb{\beta}\left[(\pmb{H}_0+\pmb{H}_2\pmb{\Psi}\pmb{H}_1)(\pmb{H}_0+\pmb{H}_2\pmb{\Psi}\pmb{H}_1)^H\right]^{-1}\right)
\end{align}
Then, $\pmb{\beta}$ and $\pmb{\Psi}$ can be iteratively optimized using the similar approaches introduced with the first simplified model, as shown in \eqref{eq:RIS-70}-\eqref{eq:RIS-73}, upon invoking the $\tilde{\pmb{F}}$ and $P_t$ provided by \eqref{eq:RIS-77} and \eqref{eq:RIS-78}. 

Built on \eqref{eq:RIS-74}-\eqref{eq:RIS-76}, \cite{8982186}
considered a weighted sum-rate maximization problem, described as
\begin{subequations}\label{eq:RIS-79}
\begin{align}
\{\pmb{F},\pmb{\Psi}\}=&\arg\max_{\pmb{F},\pmb{\Psi}}\left\{\sum_{k=1}^Kw_k\log(1+\gamma_k)\right\}\\
s.t.~~C1:&~~\sum_{k=1}^K\|\pmb{f}_k\|^2\leq P_B^{\max}\\
C2:&~~|\pmb{\Psi}(n,n)|=1,~~\forall n=1,2,\ldots,N
\end{align}
\end{subequations}
where $\{w_k\}$ are weights.  This problem can also be sloved using
the iterative method. Specifically, when given $\pmb{\Psi}$, $\pmb{F}$
can be optimized using, such as, the weighted MMSE
algorithm~\cite{5756489}. Then, for a given $\pmb{F}$, the phase
shift matrix can be optimized by the Riemannian conjugate
gradient method~\cite{8982186}
to maximize the weighted sum-rate. Additionally, in \cite{8982186},
the non-convex block coordinate descent (BCD) method~\cite{120891009}
was introduced to solve the optimization problem described by \eqref{eq:RIS-79},
in the cases when CSI is ideal or non-ideal.

Instead of using the sum-rate as the optimization objective, as above
shown, a more practical optimization problem can be designed to
minimize the total transmit power, which is described
as~\cite{8811733}
\begin{subequations}\label{eq:RIS-80}
\begin{align}
\{\pmb{F},\pmb{\Psi}\}=&\arg\min_{\pmb{F},\pmb{\Psi}}\left\{\sum_{k=1}^K\left\|\pmb{f}_k\right\|^2\right\}\\
s.t.&~~\gamma_k\geq\gamma_k^{\min},~\forall k\\
&~~|\pmb{\Psi}(n,n)|=1,~~\forall n=1,2,\ldots,N
\end{align}
\end{subequations}
where $\gamma_k^{\min}$ is the minimum SINR requirement of UE $k$. The
solution to this problem can be derived via the iterative
optimization of two sub-problems~\cite{8811733}.  One optimizes
$\pmb{F}$ for a fixed $\pmb{\Psi}$ using, such as, the second-order
cone program (SOCP), convex semidefinite program (SDP), etc.  After
some transform and approximation, the second one optimizes
$\pmb{\Psi}$ for the fixed $\pmb{F}$, which can be solved using the
existing convex optimization methods, such as, CVX~\cite{cvx}.

Let us continue on the second simplified model by assuming that
$M>>K$, $N>>K$, and both $M$ and $N$ are very large, i.e., in the concept
of massive MIMO. Then, from the principles of massive MIMO, the
rows of $\left(\pmb{H}_0+\pmb{H}_2\pmb{\Psi}\pmb{H}_1\right)$ in
\eqref{eq:RIS-75} are nearly orthogonal to each other, regardless of the phase shifts applied in $\pmb{\Psi}$. This in turn means that
UEs can be spatially separated without interference between each
other. Hence, the matched-filtering (MF)
precoder is nearly
optimum. Consequently, the precoder $\pmb{F}$ and $\pmb{\Psi}$ can be
firstly optimized with respect to individual UEs. Then, a phase-shift
matrix $\pmb{\Psi}$ is obtained by combining the individual
phase-shift matrices obtained from the individual optimizations. Below
an exemplified method towards the objective is
analyzed.

Since it is in the principle of massive MIMO, we can assume that the
DPL and RPL of a UE are orthogonal. Hence, the precoders for DPL and
RPL can be optimized separately without any loss of performance,
provided that the signals conveyed by DPL and RPL are coherently added
at a UE. Considering this property and expressing $\pmb{\Psi}^{(k)}=
\pmb{\Psi}_2^{(k)}\pmb{\Psi}_1^{(k)}$, where
$\pmb{\Psi}_i^{(k)}=\textrm{diag}\{e^{j\psi_{i1}^{(k)}},e^{j\psi_{i2}^{(k)}},\ldots,e^{j\psi_{iN}^{(k)}}\}$
for $i=1,2$, the observation at UE $k$ can be represented as
\begin{align}\label{eq:RIS-83}
y_k=&\left(\sqrt{\rho_k}\pmb{h}_{k,0}^T\pmb{f}_{k,D}+\sqrt{1-\rho_k}\pmb{h}_{k,2}^T\pmb{\Psi}_2^{(k)}\pmb{\Psi}_1^{(k)}\pmb{H}_1\pmb{f}_{k,R}\right)x_k+n_k,\nonumber\\
&~k=1,2,\ldots,K
\end{align}
where, again, $\rho_k$ implements the power-allocation between the DPL
and RPL of UE $k$. When $a_k=|\pmb{h}_{k,0}^T\pmb{f}_{k,D}|$ and
$b_k=|\pmb{h}_{k,2}^T\pmb{\Psi}_2^{(k)}\pmb{\Psi}_1^{(k)}\pmb{H}_1\pmb{f}_{k,R}|$
are known, $\rho_k$ can be found from \eqref{eq:RIS-42}.

From \eqref{eq:RIS-83}, the optimum precoder for DPL is
\begin{align}\label{eq:RIS-84}
\pmb{f}_{k,D}=\frac{\pmb{h}_{k,0}^*}{\sqrt{\|\pmb{h}_{k,0}\|^2}}
\end{align}
Given $\pmb{\Psi}_1^{(k)}$ and $\pmb{\Psi}_2^{(k)}$, the optimum precoder for RPL is
\begin{align}\label{eq:RIS-85}
\pmb{f}_{k,R}=\frac{\pmb{H}_1^H\left(\pmb{\Psi}_1^{(k)}\right)^*\left(\pmb{\Psi}_2^{(k)}\right)^*\pmb{h}_{k,2}^*}{\sqrt{\left\|\pmb{h}_{k,2}^T\pmb{\Psi}_2^{(k)}\pmb{\Psi}_1^{(k)}\pmb{H}_1\right\|^2}}
\end{align}
which is a function of $\pmb{\Psi}_1^{(k)}$ and $\pmb{\Psi}_2^{(k)}$
that need to be determined.

Alternatively, when $\pmb{f}_{k,R}$ is given, then, let express
$\pmb{H}_1=\left[\pmb{h}_{1,1},\pmb{h}_{1,2},\ldots,\pmb{h}_{1,N}\right]^T$,
where $\pmb{h}_{1,n}^T\in\mathbb{C}^{1\times M}$ denotes the $n$th row
of $\pmb{H}_1$, the reflected term in \eqref{eq:RIS-83} can be
expressed as
\begin{align}\label{eq:RIS-86}
b_k=&\pmb{h}_{k,2}^T\pmb{\Psi}_2^{(k)}\pmb{\Psi}_1^{(k)}\pmb{H}_1\pmb{f}_{k,R}\nonumber\\
=&\sum_{n=1}^N\left(h_{k,2,n}e^{j\psi_{2n}^{(k)}}\right)\times \left(e^{j\psi_{1n}^{(k)}}\pmb{h}_{1,n}^T\pmb{f}_{k,R}\right)
\end{align}
Explicitly, to add coherently the signals arriving at UE $k$, the
phase shifts of RIS have the determinate solutions of
\begin{align}\label{eq:RIS-87}
\psi_{2n}^{(k)}=&-\arg\left(h_{k,2,n}\right)\nonumber\\
\psi_{1n}^{(k)}=&-\arg\left(\pmb{h}_{1,n}^T\pmb{f}_{k,R}\right)
\end{align}
for $n=1,2,\ldots,N$, where $\arg(x)$ returns the phase of the complex
number $x$. Consequently, \eqref{eq:RIS-86} can be expressed as
\begin{align}\label{eq:RIS-88}
b_k=&\sum_{n=1}^N|h_{k,2,n}||\pmb{h}_{1,n}^T\pmb{f}_{k,R}|
\end{align}
What left here is to derive a $\pmb{f}_{k,R}$ from the optimization problem
\begin{subequations}\label{eq:RIS-89}
\begin{align}
\pmb{f}_{k,R}^{opt}=&\arg\max_{\pmb{f}_{k,R}}\left\{\sum_{n=1}^N|h_{k,2,n}||\pmb{h}_{1,n}^T\pmb{f}_{k,R}|\right\}\\
s.t.&~\|\pmb{f}_{k,R}\|^2=1
\end{align}
\end{subequations}

Inferred by the principles of massive MIMO, the solution to
$\pmb{f}_{k,R}^{opt}$ should be in the form of
\begin{align}\label{eq:RIS-90}
\pmb{f}_{k,R}^{opt}=\frac{1}{Q}\sum_{n=1}^N\omega_n\pmb{h}_{1,n}^*
\end{align}
where $Q$ is for normalizing $\pmb{f}^{opt}_{k,R}$ to unit length,
$\{\omega_n\}$ are the weights needing optimization under the
constraint of $\sum_{n=1}^N\omega_n=1$. Substituting \eqref{eq:RIS-90}
into \eqref{eq:RIS-89} and making use of the property of
$\pmb{h}_{1,i}^T\pmb{h}_{1,j}^*\approx 0$, $\forall i\neq j$ when $M$
is large, we obtain the objective function
\begin{align}\label{eq:RIS-91}
J(\{\omega_i\})=\sum_{i=1}^N\omega_i\lambda_i
\end{align}
where $\lambda_i=|h_{k,2,i}||\pmb{h}_{1,i}|^2$, which
are given values. Therefore, $\{\omega_i\}$ can be obtained from the solutions of the optimization problem described as
\begin{subequations}\label{eq:RIS-92}
\begin{align}
\{\omega_n^{opt}\}=&\arg\max_{\{\omega_n\}}\left\{J(\{\omega_n\})=\sum_{n=1}^N\omega_n\lambda_n\right\}\\
s.t. &~\sum_{n=1}^N\omega_n=1\\
&~\omega_n\geq 0, \forall n
\end{align}
\end{subequations}
This is a simple linear programming
problem~\cite{book-Antoniou-2007,book-Andrzej-Ruszczynski-2006,book-Andreas-Practical-Optimization}. The
solution is $\omega_{i_{opt}}=1$ assigned to the
$\lambda_{i_{opt}}=\max\{\lambda_n\}$, yielding
\begin{align}\label{eq:RIS-93}
\pmb{f}_{k,R}^{opt}=\frac{\pmb{h}_{1,{i_{opt}}}^*}{\sqrt{\|\pmb{h}_{1,i_{opt}}\|^2}}
\end{align}

When the RIS-mmWave system supports $K=1$ UE, the design of precoder
and phase-shifts finishes at this point. However, the current
RIS-mmWave system supports $K>1$ UEs. For each of UEs, a separate
$\pmb{\Psi}^{(k)}$ is generated, and the $\pmb{\Psi}^{(k)}$'s of
different UEs are different. Hence, after obtaining
$\pmb{\Psi}^{(k)}=\pmb{\Psi}^{(k)}_2\pmb{\Psi}^{(k)}_1$ corresponding
to the $K$ individual UEs, a joint $\pmb{\Psi}$ needs to be designed,
which can be conceptually formulated as
\begin{subequations}\label{eq:RIS-94}
\begin{align}
\pmb{\Psi}=&\arg_ff\left[\pmb{\Psi}^{(1)},\pmb{\Psi}^{(2)},\ldots,\pmb{\Psi}^{(K)}\right]\\
s.t. &~|\pmb{\Psi}(n,n)|=1,\forall n=1,2,\ldots,N
\end{align}
\end{subequations}
where $\arg_ff\left[\cdots\right]$ means to find a function for the
purpose. Below are some possible approaches for the objective. First, with the
aid of the properties of massive MIMO, which implies that the channels
of individual UEs are similar and nearly orthogonal, a reflection
matrix can be obtained as
\begin{align}\label{eq:RIS-95}
\pmb{\Psi}=&\pmb{\Psi}_c^{-1}\sum_{k=1}^K\pmb{\Psi}^{(k)}=\pmb{\Psi}_c^{-1}\sum_{k=1}^K\pmb{\Psi}_2^{(k)}\pmb{\Psi}_1^{(k)}
\end{align}
where $\pmb{\Psi}_c$ is a $(N\times N)$ diagonal matrix containing the
real values to achieve the power constraint on RIS elements or the constraint
of $|\pmb{\Psi}(n,n)|=1$. Specifically, if RIS is able to implement
digital reflection, i.e., it can amplify the signals incident on the
surface, we can let $\pmb{\Psi}_c=\Psi_c\pmb{I}_N$, where the constant
$\Psi_c$ is chosen as
\begin{align}\label{eq:RIS-96}
\Psi_c=\sqrt{\frac{P_R}{\textrm{Tr}\left(\left[\sum_{k=1}^K\pmb{\Psi}^{(k)}\right]\left[\sum_{k=1}^K\pmb{\Psi}^{(k)}\right]^*\right)}}
\end{align}
In this case, the reflection matrices are added linearly, enabling
that a UE's signals sent by BS and impinging on RIS surface as well as
the signals conveyed on DPL and RPL are coherently added at the
UE. Consequently, if the massive RIS-mmWave system makes all UEs'
channels orthogonal with each other, the design of precoders and
reflection matrices completes. Otherwise, a further updating of
precoders in the principles of, such as ZF, may be implemented to
remove the inter-interference between UEs.

By contrast, if the RIS reflectors are fully passive reflectors
requiring $|\pmb{\Psi}(n,n)|=1$, the value of $\pmb{\Psi}_c(n,n)$ is
then given by
\begin{align}\label{eq:RIS-97}
\pmb{\Psi}_c(n,n)=|\sum_{k=1}^Ke^{j\psi_n^{(k)}}|
\end{align}
In this case, the reflection matrices are added non-linearly, i.e.,
different diagonal elements in a reflection matrix may be weighted by
different values given by \eqref{eq:RIS-97}. Consequently, the joint
reflection matrix \eqref{eq:RIS-95} is highly likely not the linear
combination of the individual reflection matrices for $K$ UEs. This
results in that a UE's signals sent by BS and impinging on RIS surface
as well as that sent on DPL are not coherently added at the UE. To
solve this problem, the BS precoders for individual UEs can be updated
using the joint phase-shift matrix \eqref{eq:RIS-95} as
\begin{align}\label{eq:RIS-98}
\pmb{f}_{k,R}^{opt}=\frac{\pmb{H}_1^H\pmb{\Psi}^*\pmb{h}_{k,2}^*}{\sqrt{\left\|\pmb{h}_{k,2}^T\pmb{\Psi}\pmb{H}_1\right\|^2}},~k=1,2,\ldots,K
\end{align}
Again, if there is still interference between UEs after the above
optimizations with respect to the individual UEs, a further ZF
precoding stage may be implemented at BS to remove the interference
between UEs.

The second method to find a joint $\pmb{\Psi}$ is formulated as
\begin{align}\label{eq:RIS-95k}
\pmb{\Psi}=&\sum_{k=1}^K\pmb{I}^{(k)}_{[N/K]}\pmb{\Psi}^{(k)}=\sum_{k=1}^K\pmb{I}^{(k)}_{[N/K]}\pmb{\Psi}_2^{(k)}\pmb{\Psi}_1^{(k)}
\end{align}
where $\pmb{I}^{(k)}_{[N/K]}$ is a $(N\times N)$ diagonal matrix
obtained from $\pmb{I}_N$ with only $N/K$ non-zero elements of one, which
assign $N/K$ RIS reflection elements to UE $k$. Explicitly, fixed
assignments or the dynamic assignments satisfying a defined objective
may be implemented. This approach guarantees that $N/K$  out
of $N$ elements are optimum towards a UE, while the other $(N-N/K)$ elements'
phases are randomly set for this UE. Hence, another stage of precoder
optimization, as shown in \eqref{eq:RIS-98}, or even a further stage
of ZF precoder design are needed.

In summary, the above processes can be summarized as Algorithm~\ref{Algorithm-RIS-mmWave-SRMU-2}.
\begin{algorithm}[ht]
\caption{Algorithm for BS to optimize $\pmb{F}$ and $\pmb{\Psi}$ in massive RIS-mmWave systems.}
\label{Algorithm-RIS-mmWave-SRMU-2}

\textbf{Inputs:} $\{\pmb{H}_{k,0}, \pmb{H}_{k,1}, \pmb{H}_{k,2}\}$ \\

\textbf{Design BS precoders for DPLs:} Compute $\pmb{f}_{k,D}$ using
\eqref{eq:RIS-84} $\forall k=1,2,\ldots,K$. \\

\textbf{Design BS precoders and RIS phase-shifts for RPLs:} 
\begin{enumerate}

\item Design $\pmb{f}_{k,R}^{opt}$ based on
  \eqref{eq:RIS-88}-\eqref{eq:RIS-93} $\forall k=1,2,\ldots,K$;

\item Design $\pmb{\Psi}$ based on \eqref{eq:RIS-87} and
  \eqref{eq:RIS-95} or \eqref{eq:RIS-95k};

\item Update $\pmb{f}_{k,R}^{opt}$ using \eqref{eq:RIS-98} $\forall
  k=1,2,\ldots,K$, if full passive RIS; or

\item based on $\pmb{\Psi}$, BS jointly designs $\pmb{F}$ for all UEs
  in the principles of ZF, or other, precoding.

\end{enumerate}

\end{algorithm}

The joint RIS phase-shift matrices \eqref{eq:RIS-95} and
\eqref{eq:RIS-95k} imply that the performance in terms of one UE
degrades, as the number of UEs increases. This is the result of two
factors. First, when there are $K$ UEs, RIS has to divide its power to
reflect signals towards $K$ directions, as shown in \eqref{eq:RIS-96} and
\eqref{eq:RIS-97}. Hence, when $K$ increases, less power is delivered
by RIS to a specific UE. Second, for passive RIS, as $K$ increases,
the phase-shift matrix $\pmb{\Psi}$ becomes more random, as explained
in \eqref{eq:RIS-95} and \eqref{eq:RIS-95k}. Hence, the signals from
BS are more randomly reflected to UEs, which also results in
performance degradation. Furthermore, we might be inferred that in
passive RIS-mmWave systems, a random reflection matrix may be expected to be
near-optimum, if $K>>1$. This however needs further research to
verify.

\section{RIS-mmWave: Multiple-RISs Multiple Users}\label{section-6G-2.2.3}

%
\begin{figure}[tb]
  \begin{center}
    \includegraphics[width=0.65\linewidth]{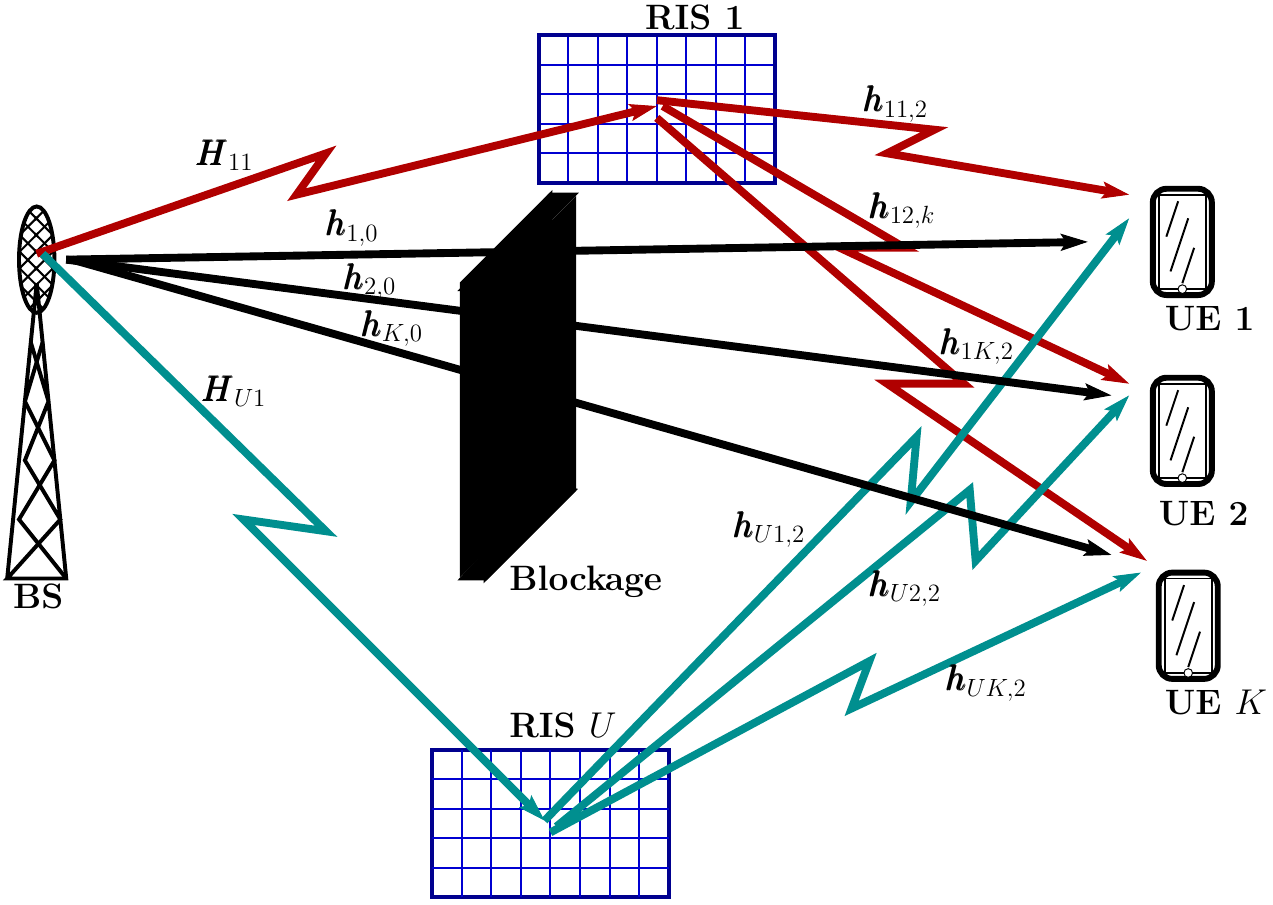} 
  \end{center}
  \caption{A RIS-mmWave system with one BS via $U$ RISs to serve $K$ user equipments.}
\label{figure-MRMU-diagram}
\end{figure}

This section further extends the RIS-mmWave model considered in
Section~\ref{section-6G-2.2.2} to a model with $U>1$
RISs~\cite{9174831,9497709}, as shown in
Fig.~\ref{figure-MRMU-diagram}. To be more specific, the model assumes
one BS with an array of $M$ antennas, $U$ RISs with the $u$th RIS
having $N_u$ reflection elements, and $K$ UEs of each equipped with
single antenna for simplicity. However, the single-antenna UEs can be
straightforwardly extended to the general UEs of each with multiple
antennas. Again, downlink transmission is considered. Under these
settings and assumptions, the baseband transmit signal of BS can be
expressed as
\begin{align}\label{eq:RIS-99}
\pmb{s}=\sum_{k=1}^K\pmb{f}_kx_k=\pmb{F}\pmb{x}
\end{align}
where the precoding vector $\pmb{f}_k\in\mathbb{C}^{M\times 1}$,
precoding matrix $\pmb{F}\in\mathbb{C}^{M\times K}$ and data vector
$\pmb{x}\in\mathbb{C}^{K\times 1}$ are the same as them in
Section~\ref{section-6G-2.2.2}.

Assume that a DPL may exist between BS and a UE, in addition to the
RPLs.  Accordingly, the received observation of UE $k$ can be
expressed as
\begin{align}\label{eq:RIS-100}
y_{k}=&\left(\pmb{h}_{k,0}^T+\sum_{u=1}^U\kappa_{uk}\pmb{h}_{uk,2}^T\pmb{\Psi}_u\pmb{H}_{u1}\right)\pmb{F}\pmb{x}+n_k,~k=1,2,\ldots,K
\end{align}
where $\kappa_{uk}$ is an association parameter, with $\kappa_{uk}=1$
denoting that the signal to UE $k$ is via RIS $u$, and $\kappa_{uk}=0$
otherwise, while the other arguments in \eqref{eq:RIS-100} have the
same definitions as those in \eqref{eq:RIS-74}, except replacing $N$
by $N_u$. Similarly, corresponding to \eqref{eq:RIS-75}, the
observations of $K$ UEs can be expressed as
\begin{align}\label{eq:RIS-101}
\pmb{y}=&\left(\pmb{H}_0+\sum_{u=1}^U\pmb{\kappa}_{u}\pmb{H}_{u,2}\pmb{\Psi}_u\pmb{H}_{u1}\right)\pmb{F}\pmb{x}+\pmb{n}
\end{align}
where
$\pmb{\kappa}=\textrm{diag}\left\{\kappa_{u1},\kappa_{u2},\ldots,\kappa_{uK}\right\}$ and $\pmb{H}_{u,2}=\left[\pmb{h}_{u1,2},\pmb{h}_{u2,2},\cdots,\pmb{h}_{uK,2}\right]^T$.

From \eqref{eq:RIS-100}, the SINR of UE $k$ is 
\begin{align}\label{eq:RIS-102}
\gamma_k=\frac{\left|\left(\pmb{h}_{k,0}^T+\sum_{u=1}^U\kappa_{uk}\pmb{h}_{uk,2}^T\pmb{\Psi}_u\pmb{H}_{u1}\right)\pmb{f}_k\right|^2}{\sum_{l\neq k}\left|\left(\pmb{h}_{k,0}^T+\sum_{u=1}^U\kappa_{uk}\pmb{h}_{uk,2}^T\pmb{\Psi}_u\pmb{H}_{u1}\right)\pmb{f}_l\right|^2+\sigma^2}
\end{align}

According to \eqref{eq:RIS-100} and \eqref{eq:RIS-101}, when given the
channels $\pmb{H}_0$, $\pmb{H}_1$ and $\{\pmb{H}_{u,2}\}$, the
optimization of the considered RIS-mmWave system includes:
\begin{itemize}

\item Associations of UEs with RISs;

\item Optimization of $\{\pmb{\Psi}_{u,2}\}$;

\item Optimization of precoder $\pmb{F}$.

\end{itemize}
Due to the high complexity and non-convexity of the joint optimization
problem, the above-mentioned optimization may only be executed separately and in
the iteration ways. For example, UE associations can be firstly carried
out based on the strength of the pilot signals sent by BS, reflected
by RISs and received by UEs. After associations, BS can choose to send
information to UEs via their associated RISs with the aid of
beamforming. Accordingly, RISs' reflection matrices
$\{\pmb{\Psi}_{u,2}\}$ and BS' precoder $\pmb{F}$ can be optimized
using the iterative methods similar to that considered in
Section~\ref{section-6G-2.2.2}.

Instead of UE association, in \cite{9497709}, switching `on-off' of
RISs is considered to improve energy-efficiency. Accordingly,
\eqref{eq:RIS-100} is represented as
\begin{align}\label{eq:RIS-103}
y_{k}=&\left(\pmb{h}_{k,0}^T+\sum_{u=1}^U\kappa_{u}\pmb{h}_{uk,2}^T\pmb{\Psi}_u\pmb{H}_{u1}\right)\pmb{F}\pmb{x}+n_k,~k=1,2,\ldots,K
\end{align}
where $\kappa_u=1$ turns `on' a RIS, while $\kappa_u=0$ switches off a
RIS. Correspondingly, the SINR of UE $k$ is
\begin{align}\label{eq:RIS-104}
\gamma_k=\frac{\left|\left(\pmb{h}_{k,0}^T+\sum_{u=1}^U\kappa_{u}\pmb{h}_{uk,2}^T\pmb{\Psi}_u\pmb{H}_{u1}\right)\pmb{f}_k\right|^2}{\displaystyle\sum_{l\neq k}\left|\left(\pmb{h}_{k,0}^T+\sum_{u=1}^U\kappa_{u}\pmb{h}_{uk,2}^T\pmb{\Psi}_u\pmb{H}_{u1}\right)\pmb{f}_l\right|^2+\sigma^2},~k=1,2,\ldots,K
\end{align}
The system's sum rate is expressed as
\begin{align}\label{eq:RIS-105}
R=B\sum_{k=1}^K\log_2\left(1+\gamma_k\right)
\end{align}
where $B$ is bandwidth.

For power consumption, the BS transmit power, BS circuit power, RIS
circuit power, and RIS transmit power are included, having the
representation of~\cite{9497709}
\begin{align}\label{eq:RIS-106}
P=\sum_{k=1}^K\left\|\pmb{f}_k\right\|^2/\eta+P_{B,c}+\sum_{u=1}^U\kappa_uN_uP_{R,e}+\sum_{k=1}^KP_{k,c}
\end{align}
where, on the righthand side, the first term denotes the power
consumed by BS for signal transmission, $\eta$ is the power
amplification efficiency of BS, the second term is the circuit (or
signal processing) power of BS, the third term is the power
consumption of $U$ RISs, with $P_{R,e}$ representing the power
consumption per RIS element, and the last term is the total power
consumed by $K$ UEs to receive their information.

Then, in \cite{9497709}, the BS precoder, reflection matrices and the
`on-off' states of RISs are optimized to maximize the
energy-efficiency under the minimum rate, $R_{k,\min}$, requirement
for each of UEs and the constraint of the total BS transmit power of
$P_B^{\max}$. The optimization problem is formulated as
\begin{subequations}\label{eq:RIS-107}
\begin{align}
&\left\{\{\kappa_u^{opt}\},\{\pmb{f}_k^{opt}\},\{\pmb{\Psi}_u^{opt}\}\right\}=\arg\max_{\{\kappa_u\},\{\pmb{f}_k\},\{\pmb{\Psi}_u}\left\{\frac{R}{P}\right\}\\
&~~~~~~~~s.t.~B\log_2\left(1+\gamma_k\right)\geq R_{k,\min},\forall k=1,2,\ldots,K\\
&~~~~~~~~~\sum_{k=1}^K\left\|\pmb{f}_k\right\|^2\leq P_B^{\max}\\
&~~~~~~~~~\kappa_u\in\{0,1\}, \forall u=1,2,\ldots,U\\
&~~~~~~~~~|\pmb{\Psi}_u(n,n)|=1, \forall n=1,\ldots,N_u;~\textrm{and~}\forall u=1,2,\ldots,U
\end{align}
\end{subequations}
The optimization problem stated in \eqref{eq:RIS-107} is a
mixed-integer nonlinear problem, which is extremely hard to solve. In
\cite{9497709}, considering the single- and multi-user cases, two
iterative algorithms were developed for solving the problem to
obtain the sub-optimum solutions.

Below we consider the massive RIS-mmWave scenario where $M>>K$ and
$N_u$ of the $u$th RIS is significantly larger than the number of UEs
associated with it. Hence, following the principles of massive MIMO,
the channels from BS to different UEs and that from BS to the same UE
via different links can be assumed to be orthogonal with each
other. We assume that the associations of UEs are carried out before
the optimization of BS precoder and RISs' reflection matrices. In
practice, this can be achieved during the beam search stage, with the
aid of the pilot beams sent by BS via all possible links and also all
possible beams~\cite{10071555}. Based on the measurements of pilot
beams, which may include signal strength, delay, AOA, AOD, and even
the positions of UE and RISs, a UE decides which RISs it associates.

Specifically, if each UE is regulated to be associated with only one RIS,
after the association, the optimization of $\pmb{F}$ and
$\{\pmb{\Psi}_u\}$ is the same as that considered in the massive
RIS-mmWave case in Section~\ref{section-6G-2.2.2}. 

By contrast, when each UE is allowed to be associated with multiple
RISs, the optimizations of $\pmb{f}_k$ and $\{\pmb{\Psi}_u^{(k)}\}$ can be
based on the observation
\begin{align}\label{eq:RIS-108}
y_{k}=&\left(\sqrt{\rho_0}\pmb{h}_{k,0}^T\pmb{f}_{k,D}+\sum_{u=1}^{U_k}\sqrt{\rho_u}\pmb{h}_{uk,2}^T\pmb{\Psi}_u^{(k)}\pmb{H}_{u1}\pmb{f}_{ku,R}\right)x_k+n_k,\nonumber\\
&~k=1,2,\ldots,K
\end{align}
where $U_k$ is the number of RISs that UE $k$ associates with. Eq.~\ref{eq:RIS-108} is the simplified equation of \eqref{eq:RIS-105} by applying the
assumptions that the channels from BS to different UEs are orthogonal,
and that the channels from BS to a UE via different links are also
orthogonal. Hence, we have 
\begin{align}\label{eq:RIS-109}
\pmb{f}_k=\sqrt{\rho_0}\pmb{f}_{k,D}+\sum_{u=1}^{U_k}\sqrt{\rho_u}\pmb{f}_{ku,R}
\end{align}
where $\pmb{f}_{k,D}$ and $\{\pmb{f}_{ku,R}\}$ are orthogonal with
each other and are unity length, $\{\rho_u\}$ are power-allocation coefficients.

From \eqref{eq:RIS-108}, we can readily know that 
\begin{subequations}\label{eq:RIS-110}
\begin{align}
\pmb{f}_{k,D}=&\frac{\pmb{h}_{k,0}^*}{\sqrt{\|\pmb{h}_{k,0}\|^2}}\\
\pmb{f}_{ku,R}=&\frac{\pmb{H}_{u1}^H\left(\pmb{\Psi}_u^{(k)}\right)^*\pmb{h}_{uk,2}^*}{\sqrt{\|\pmb{h}_{uk,2}^T\pmb{\Psi}_u^{(k)}\pmb{H}_{u1}\|^2}},~\forall~u=1,2,\ldots,U_k
\end{align}
\end{subequations}
Substituting \eqref{eq:RIS-110} into \eqref{eq:RIS-108} gives
\begin{align}\label{eq:RIS-112}
y_{k}=&\left(\sqrt{\rho_0}\sqrt{\|\pmb{h}_{k,0}\|^2}+\sum_{u=1}^{U_k}\sqrt{\rho_u}\sqrt{\|\pmb{h}_{uk,2}^T\pmb{\Psi}_u^{(k)}\pmb{H}_{u1}\|^2}\right)x_k+n_k
\end{align}
where $\pmb{\Psi}_u^{(k)}$ can be designed to maximize $\|\pmb{h}_{uk,2}^T\pmb{\Psi}_u^{(k)}\pmb{H}_{u1}\|^2$~\cite{9497709}. Then, a $\pmb{\Psi}_u$ that is common to all UEs associated with the $u$th RIS is calculated and $\pmb{f}_k$ may be further updated by following the discussion in Section~\ref{section-6G-2.2.2}.

From \eqref{eq:RIS-112}, the SNR of UE $k$ is
\begin{align}\label{eq:RIS-113}
\gamma_{k}=&\frac{\left(\sqrt{\rho_0}\sqrt{\|\pmb{h}_{k,0}\|^2}+\sum_{u=1}^{U_k}\sqrt{\rho_u}\sqrt{\|\pmb{h}_{uk,2}^T\pmb{\Psi}_u\pmb{H}_{u1}\|^2}\right)^2}{\sigma^2}
\end{align}
which can be maximized via power-allocation. Let $\lambda_0=\sqrt{\|\pmb{h}_{k,0}\|^2}$ and $\lambda_u=\sqrt{\|\pmb{h}_{uk,2}^T\pmb{\Psi}_u\pmb{H}_{u1}\|^2}$. Then, \eqref{eq:RIS-113} is represented as
\begin{align}\label{eq:RIS-114}
\gamma_{k}=&\frac{\left(\sum_{u=0}^{U_k}\sqrt{\rho_u}\lambda_u\right)^2}{\sigma^2}
\end{align}
As $\{\lambda_u\}$ are positive numbers, the optimization problem for obtaining $\{\rho_u\}$ can be stated as
\begin{subequations}\label{eq:RIS-115}
\begin{align}
\{\rho_u^{opt}\}&=\arg\max_{\{\rho_u\}}\left\{\sum_{u=0}^{U_k}\sqrt{\rho_u}\lambda_u\right\}\\
s.t.&~~\sum_{u=0}^{U_k}\rho_u=1
\end{align}
\end{subequations}
which can be readily solved using the method of Lagrange multipliers, yielding
\begin{align}\label{eq:RIS-116}
\rho_u=\frac{\lambda_u^2}{\sum_{l=0}^{U_k}\lambda_l^2},~u=0,1,\ldots,U_k
\end{align}
Substituting them to \eqref{eq:RIS-114} obtains
\begin{align}\label{eq:RIS-117}
\gamma_{k}=&\frac{\sum_{u=0}^{U_k}\lambda_u^2}{\sigma^2}\nonumber\\
=&\frac{\|\pmb{h}_{k,0}\|^2+\sum_{u=1}^{U_k}\|\pmb{h}_{uk,2}^T\pmb{\Psi}_u\pmb{H}_{u1}\|^2}{\sigma^2}
\end{align}
showing that the power-allocation by BS to the beams towards one UE implements the MRC result at the UE. Following the principle of MRC, higher power is assigned to a stronger beam. If a beam is weak, BS assigns it little power. In practice, the weak beams may be simply switched off to save power of both BS and RISs, also because the channel estimation of weak beams is difficult and unreliable.     


\section{Concluding Remarks}\label{section-6G-2.3}

This chapter addressed the principles of RIS-mmWave and the
optimization in RIS-mmWave systems. First, a conceptual example was
provided to explain the various wireless communication implementation
strategies, when receiver, transmitter or/and channel reflector(s) have
CSI. Then, the optimization in the context of three RIS-mmWave models
were respectively analyzed. All the three models assume one BS, while
the first one assumes one RIS and one UE, the second one assumes one
RIS and multiple UEs, and the third one assumes multiple RISs and
multiple UEs. Furthermore, some sub-models under the above three
were also considered. The analyses show that in the simplest model
with one UE, the optimization objective is to design BS precoder and
RIS phase-shifts, so that the signals arriving at UE are coherently added.
By contrast, when there are multiple UEs, the optimization needs to
mitigate inter-UE interference, in addition to maximizing individual
UE's performance. Furthermore, when there are multiple RISs, a stage
of optimization for UEs' association with RISs is required. Additionally,
it is shown that when RIS-mmWave systems satisfy the massive MIMO
conditions, the BS precoder and RIS phase-shifts of individual UEs can
be independently designed, making the optimization relatively simple.

While the deployment of RISs can help solve the blockage problem in
mmWave communications, there are also challenges in the design and
optimization for high-efficiency. The first and also most critical
challenge is channel acquisition. As shown in
Sections~\ref{section-6G-2.2.1}-\ref{section-6G-2.2.3}, in a
RIS-mmWave system implementing CSI-relied RIS optimization, there are
three channels, namely $\pmb{H}_0$, $\pmb{H}_1$ and $\pmb{H}_2$, to
estimate. While $\pmb{H}_0$ of DPL can be estimated by the
conventional methods, the acquisitions of $\pmb{H}_1$ and $\pmb{H}_2$
of RPL are not straightforward. This becomes especially difficult, when
a RIS is equipped with a large number of reflection elements, which,
however, may be the case in practice. Here, the main challenge is that
RIS is typically assumed to be simple, passive operation, and of
extremely low-power consumption. Furthermore, the size of individual
reflection elements is supposed to be small, which leads to the channel acquisition with weak signals. Because of the above-mentioned
constraints plus channel acquisition under weak signals, it is
expected that RIS can hardly achieve satisfactory channel acquisition
by itself.

Another challenge is the RIS phase-shift optimization, especially,
when the number reflection elements is large. From the analyses in
Sections~\ref{section-6G-2.2.1}-\ref{section-6G-2.2.3} we can realize
that whenever the CSI-relied phase-shift optimization is implemented,
the optimization of RIS-mmWave systems becomes more involved and
challenging, regardless of the active RIS or passive RIS
employed. Furthermore, it seems that when the number of UEs increases,
the benefit added by the CSI-relied RIS optimization reduces, which is
especially true, when passive RIS is assumed due to the unit amplitude
constraint on per reflection element. For example, assume a three UE
system where a RIS phase-shift is required to set $e^{j\pi/6}$ for UE
1, $e^{j\pi/3}$ for UE 2, and $e^{j\pi/2}$ for UE 3. As the
phase-shift can only be set to one value, by adding these values and
normalizing the result to unit amplitude, the actual phase shift
configured is about $e^{j\pi/3}$ and the amplitude normalization factor
is about $2.73$. When another reflection element has three
corresponding values of $e^{j\pi/6}$, $e^{j\pi/4}$ and $e^{j\pi/3}$,
its phase shift is $e^{j\pi/4}$ and the amplitude normalization factor
is $1.47$. This simple example explains that in passive RIS, the
optimum phase-shift matrices designed from the optimization with respect to
individual UEs may not be beneficial in general, when the number of
UEs is big. However, the situation in massive RIS-mmWave expects further research. 

The solutions around the above dilemma might be as follows. First, the
RIS phase-shift matrix $\pmb{\Psi}$ may simply be randomly set. In
this case, only the composite channel $\pmb{H}_2\pmb{\Psi}\pmb{H}_1$
needs to be estimated, which significantly lessens the challenges from
channel estimation and system optimization. The second approach is
discretizing $\pmb{\Psi}$ into a set $\{\pmb{\Psi}_i\}$, which may be
designed according to the specific application scenarios. In this
case, BS only needs to estimate the composite channels associate with
these candidate phase-shift matrices during optimization. Finally, in
the massive RIS-mmWave scenarios, the method suggested by \eqref{eq:RIS-95k} might be practical, but further research for
confirmation is expected.


\end{document}